\documentclass[aps, prx, twocolumn, superscriptaddress, longbibliography]{revtex4-2}

\usepackage{graphicx}
\usepackage{amsmath}
\usepackage{amssymb}
\usepackage{amsthm}
\usepackage{bbold}
\usepackage{pifont}
\usepackage{array}
\usepackage{tabularx}
\usepackage{hyperref}
\hypersetup{
 pdftitle = {Correlations enable lossless ergotropy transport},
 pdfauthor = {R. Simon, J. Anders, K. Hovhannisyan},
 breaklinks=true,
 pdfnewwindow=true,
 colorlinks=true,
 linkcolor=red,
 citecolor=blue,
 filecolor=blue,
 urlcolor=blue
}
\usepackage{xcolor}
\usepackage{psfrag}
\usepackage{soul}
\usepackage{textcomp}
\usepackage{bm}
\usepackage{bbm}
\usepackage{yfonts}

\usepackage[utf8]{inputenc}
\usepackage[english]{babel}
\usepackage[T1]{fontenc}
\usepackage{dcolumn}

\usepackage{tikz}
\usetikzlibrary{quantikz}

\newtheorem{lemma}{Lemma}

\newtheoremstyle{PRLemma}
  {0pt}
  {0pt}
  {}
  {10pt}
  {\itshape}
  {\!.---}
  {0pt}
  {\thmname{#1}\thmnumber{ #2}\textnormal{\thmnote{ (#3)}}}

\theoremstyle{PRLemma}

\DeclareMathOperator{\tr}{Tr}
\DeclareMathOperator{\diag}{diag}

\DeclareMathOperator{\spec}{Spec}

\DeclareMathOperator{\prob}{Prob}

\newcommand{\comments}[1]{}

\newcommand{\av}[1]{\langle #1 \rangle}

\newcommand{\id}{\mathbbm{1}}
\newcommand{\nul}{\mathbb{0}}
\newcommand{\RN}[1]{\textup{\uppercase\expandafter{\romannumeral#1}}}

\newcommand{\erg}{\mathcal{E}}
\newcommand{\ergt}{\mathcal{E}_{\mathrm{tot}}}

\newcommand{\G}{\mathrm{G}}
\newcommand{\trho}{\tilde{\rho}}

\renewcommand{\cha}{\mathrm{ch}}
\newcommand{\dra}{\mathrm{dr}}
\newcommand{\cyc}{\mathrm{cyc}}
\newcommand{\D}{\mathcal{D}}
\newcommand{\DBu}{\mathcal{D}_{\mathrm{B}}}

\newcommand{\DHS}{\mathcal{D}_{\mathrm{HS}}}
\newcommand{\opn}[1]{\Vert #1 \Vert_{\mathrm{op}}}

\begin{document}

\title{Correlations enable lossless ergotropy transport}

\author{Rick P. A. Simon}
\email{rick.paul.axel.simon@gmail.com}
\affiliation{University of Potsdam, Institute of Physics and Astronomy, Karl-Liebknecht-Str. 24-25, 14476 Potsdam, Germany}
\affiliation{Department of Physics and Astronomy, University of Exeter, Stocker Road, Exeter EX4 4QL, UK}

\author{Janet Anders}
\affiliation{University of Potsdam, Institute of Physics and Astronomy, Karl-Liebknecht-Str. 24-25, 14476 Potsdam, Germany}
\affiliation{Department of Physics and Astronomy, University of Exeter, Stocker Road, Exeter EX4 4QL, UK}

\author{Karen V. Hovhannisyan}
\email{karen.hovhannisyan@uni-potsdam.de}
\affiliation{University of Potsdam, Institute of Physics and Astronomy, Karl-Liebknecht-Str. 24-25, 14476 Potsdam, Germany}

\begin{abstract}

``A battery powers a device'' can be read as ``work stored in the battery is being transported to the device.'' In quantum batteries, the total amount of stored work can be measured by ergotropy, which is the maximal work extractable by unitary operations. Transporting ergotropy is fundamentally different from transporting energy, and here we find that ergotropy can be \textit{gained} even when the transmission channel is strictly energy conserving. We show that, generically, ergotropy transport is lossy whenever the two systems start uncorrelated. In contrast, for a large class of correlated initial states, transport can be gainful. Furthermore, a single correlated state can be used multiple times, allowing to transport without losses an order of magnitude more work than the battery capacity. Correlations are thus a useful resource for ergotropy transport, and we quantify how this resource is consumed during gainful transport.

\end{abstract}

\maketitle

\textit{Introduction.}---Batteries store work, it is their sole function. By storing work we mean storing energy that is available to be extracted in the form of work. Many fundamental aspects of quantum batteries' storage capacity \cite{Alicki_2013, Perarnau_2015, Andolina_2019, Tirone_2022, Yang_2023}, charging speed \cite{Hovhannisyan_2013, Binder_2015, Campaioli_2017, Ferraro_2018, Gyhm_2022, Gyhm_2024}, and stability \cite{Santos_2019, Farina_2019, Barra_2019, Liu_2019, Pirmoradian_2019, Hovhannisyan_2020, Gherardini_2020, Quach_2020, Kamin_2020, Barra_2022, Song_2024} have been studied \cite{Campaioli_2023}. However, the no less fundamental matter of transporting a given amount of ``charge'' (i.e., work) from the battery to a consumption unit without external energetic intervention has so far received little attention \cite{Monsel_2020, Tirone_2021, Tirone_2022, Lobejko_2022, Song_2024}. A quantum battery autonomously powering a (quantum) device is a paradigmatic example of this scenario.

The charge level of a quantum battery, i.e., the total amount of work stored in it, is standardly quantified by ergotropy \cite{Alicki_2013, Gelbwaser-Klimovsky_2013, Hovhannisyan_2013, Perarnau_2015, Binder_2015, Niedenzu_2016, Campaioli_2017, Seah_2018, Ferraro_2018, Barra_2019, Farina_2019, Andolina_2019, Santos_2019, Niedenzu_2019, Monsel_2020, Gherardini_2020, Hovhannisyan_2020, Quach_2020, Cakmak_2020, Kamin_2020, Seah_2021, Tirone_2021, Barra_2022, Gyhm_2022, Tirone_2022, Lobejko_2022, Yang_2023, Campaioli_2023, Song_2024}. It is the maximal amount of work extractable from a system \textit{unitarily} \cite{Allahverdyan_2004}. However, a battery performing work on a consumption unit means that the two systems interact with each other. And when they are of comparable size, the process will generally not be unitary for either system. This makes the problem of ergotropy transport highly nontrivial: When will this nonunitarity cause losses during transport? Can nonunitarity be leveraged to facilitate transport?

In this Letter, we put forth a rigorous framework for autonomous ergotropy transport between two quantum systems. We establish that the efficiency of transport ultimately depends on the initial correlations between the battery and the consumption unit. If they start uncorrelated, ergotropy will generically be partially lost during transport. However, if they start correlated, the transport may even be gainful, i.e., the consumption unit may receive more ergotropy than the battery sends.

\medskip

\textit{Ergotropy and its transport.}---If $H$ is the Hamiltonian of a system that lives in a $d$-dimensional Hilbert
space, and $\rho$ is its state, then the ergotropy is \cite{Allahverdyan_2004}
\begin{align} \label{ergot_def_1}
    \erg(\rho) := \tr(\rho H) - \min\limits_U \tr(U \rho U^\dagger H).
\end{align}
Here the minimization is over \textit{all} unitary operations.

\begin{figure}[t!]
    \centering
    \includegraphics[width=0.95\columnwidth]{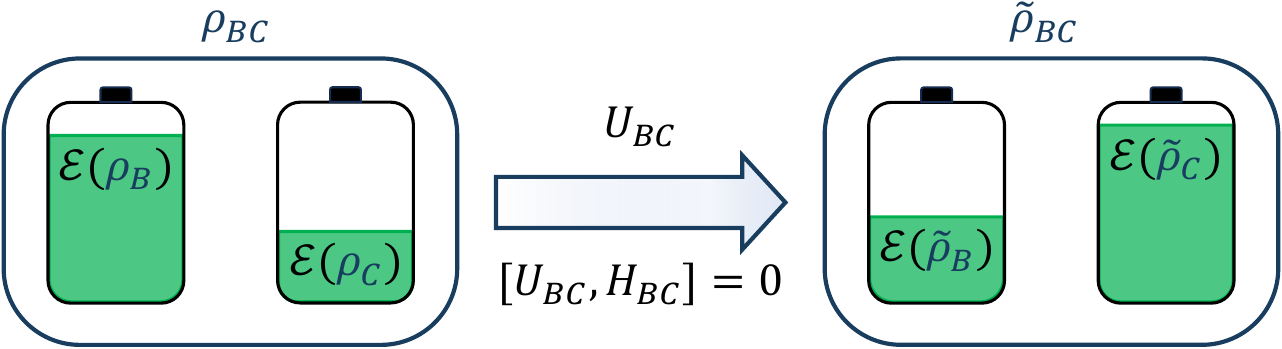}
    \caption{Schematic of transporting ergotropy from battery ($B$) to consumption unit ($C$). Initially, the systems do not interact, and their ergotropies are $\erg(\rho_B)$ and $\erg(\rho_C)$. Then an energy-conserving unitary $U_{BC}$ transforms the initial joint state $\rho_{BC}$ into $\tilde{\rho}_{BC}$, changing local ergotropies to $\erg(\tilde{\rho}_B)$ and $\erg(\tilde{\rho}_C)$. If $C$ gains more ergotropy than $B$ loses, then the transport is gainful. Otherwise, it is lossy. One of our main results is that the amount of correlations between $B$ and $C$
    is a key resource for achieving a gainful transport.}
    \label{fig:setup}
\end{figure}

In this Letter, we focus on the bipartite scenario of transporting ergotropy from system $B$ (battery) to system $C$ (consumption unit).
To make the bookkeeping of energy and entropy as clean as possible, we choose the transport ``channel'' to be strictly energy conserving
and unitary. Namely, the operation that acts on $BC$ to perform the transport is unitary, call it $U_{BC}$, and strict energy conservation means \cite{Partovi_1989, Brailovskii_1996, Janzing_2000, Brandao_2013, Chiribella_2017}
\begin{align} \label{erg_con_uni}
    [U_{BC}, H_{BC}] = \nul,
\end{align}
where $\nul$ is the zero operator and
\begin{align} \label{totham}
    H_{BC} = H_B \otimes \id_C + \id_B \otimes H_C
\end{align}
is the total Hamiltonian of $BC$ before and after the transport (during the transport, the Hamiltonian is varied in time so as to generate $U_{BC}$). Here $H_B$ and $H_C$ are, respectively, the Hamiltonians of $B$ and $C$, and $\id$ is the identity operator. This process is pictured in Fig.~\ref{fig:setup}.

The local evolution on $B$ and $C$ induced by $U_{BC}$ can be unitary only under very restrictive conditions \cite{Brandao_2013, Aberg_2014, Vaccaro_2018, Woods_2019}. In general, the individual evolutions of $B$ and $C$ will be (highly) nonunitary. Due to this, ergotropy transport is fundamentally different from energy transport \cite{Partovi_2008, Jennings_2010, Jevtic_2012, Henao_2018, Andolina_2018, Lipka-Bartosik_2024} as well as the process of increasing one system's average energy at the expense of another system's ergotropy \cite{Lobejko_2020, Biswas_2022}. For example, when a thermal system is heated up, its energy increases, but its ergotropy remains zero. Conversely, it is possible to purify a system's state without changing its energy, which means its ergotropy increases while its energy remains fixed.


\medskip

The main figure of merit we consider is the quality of the transport. It is naturally characterized by how much the channel adds to the ergotropy it transports. We call the added amount \textit{ergotropy gain}. By definition, it is the difference between the amount of ergotropy lost by $B$ and that received by $C$:
\begin{align} \label{erggain}
    \erg_\G := \erg(\trho_B) + \erg(\trho_C) - \erg(\rho_B) - \erg(\rho_C),
\end{align}
where $\rho_B = \tr_C[\rho_{BC}]$ and $\rho_C = \tr_B[\rho_{BC}]$ are the marginals of the initial state of $BC$, $\rho_{BC}$. Similarly, $\trho_B$ and $\trho_C$ are the marginals of $\trho_{BC} = U_{BC} \rho_{BC} U_{BC}^\dagger$, which is the final, post-transport state of $BC$. When $\erg_\G < 0$, namely, when ergotropy is dissipated during transport, we say that the channel is lossy. When $\erg_\G \geq 0$, it is lossless, and gainful when $\erg_\G > 0$.

As Eq.~\eqref{erggain} suggests, the ergotropy gain can also be interpreted as the difference between final and initial values of the ``locally extractable ergotropy.'' The latter is the maximal work extractable by unitary operations of the form $U_B \otimes U_C$ and is obviously equal to $\erg(\rho_B) + \erg(\rho_C)$. Here ``local'' refers to quantities pertaining to individual systems ($B$ or $C$), to contrast with $BC$-wide (``global'') quantities. This observation allows us to connect the ergotropy gain to the so-called ``ergotropic gap'' \cite{Mukherjee_2016}, defined as the difference between the global and locally extractable ergotropies:
\begin{align}
    \delta^\erg_{BC} := \erg(\rho_{BC}) - \erg(\rho_B) - \erg(\rho_C) \geq 0.
\end{align}
The ergotropic gap, $\delta^\erg_{BC}$, gives the amount of work hidden from local operations.

Now, as directly follows from Eq.~\eqref{ergot_def_1}, the action of an energy-conserving unitary on a state does not change its ergotropy. Hence, since $U_{BC}$ is energy conserving, $\erg(\trho_{BC}) = \erg(\rho_{BC})$. Therefore, the ergotropy gain in Eq.~\eqref{erggain} can be cast as a change in the ergotropic gap:
\begin{align} \label{gain_vs_gap}
    \erg_\G = \delta^\erg_{BC} - \tilde{\delta}^\erg_{BC}.
\end{align}
Thus, positive ergotropy gain is achieved when the energy-conserving unitary $U_{BC}$ ``redistributes'' the ergotropy in the total system so that more of the global ergotropy becomes locally extractable. The resource expended for achieving the gain is the ergotropic gap. Note that $U_{BC}$ acts globally, and therefore $\tilde{\rho}_X$ is generally not reachable from $\rho_X$ by local unitary transformations.

An immediate consequence of Eq.~\eqref{gain_vs_gap} and the fact that the ergotropic gap is a nonnegative quantity is the following lemma.
\begin{lemma} \label{thm:lossy}
Ergotropy transport cannot be gainful, i.e., one always has $\erg_\G \leq 0$, whenever $\delta^\erg_{BC} = 0$.
\end{lemma}

Thus, for gainful transport to be possible, the initial ergotropic gap must be positive. Importantly, the presence of a nonzero ergotropic gap is linked to the correlations in the system \cite{Perarnau_2015, Mukherjee_2016, Alimuddin_2019, Sen_2021, Puliyil_2022}. It can be used to witness entanglement \cite{Perarnau_2015, Alimuddin_2019, Sen_2021, Puliyil_2022} or quantum discord \cite{Mukherjee_2016}. In general, however, the relation between ergotropic gap and correlations is not exact: there exist uncorrelated states for which $\delta^\erg_{BC} > 0$ and correlated states for which $\delta^\erg_{BC} = 0$. An example of the former situation is the phenomenon of activation, where two uncorrelated copies of a passive state $\rho$ turn out non-passive (``active'') when considered jointly \cite{Pusz_1978, Lenard_1978}.

Activation can happen only when $d_{BC} > 4$. When $d_{BC} = 4$, namely, each system is a qubit, all passive states are Gibbs states (i.e., are $\propto e^{-\beta H}$, $\beta > 0$), so no activation is possible \cite{Pusz_1978, Lenard_1978}. In fact, the following lemma shows that \textit{any} two-qubit product state leads to non-gainful ergotropy transport.

\begin{lemma} \label{thm:2qbt}
For two qubits, one has $\erg_\G \leq 0$ whenever the initial state is uncorrelated; i.e., $\rho_{BC} = \rho_B \otimes \rho_C$.
\end{lemma}
Importantly, Lemma~\ref{thm:2qbt} does not imply a ``deterministic'' relation between $\erg_\G$ and correlations---their presence is not sufficient for $\erg_\G > 0$. For a pertinent example, as well as a proof of Lemma~\ref{thm:2qbt}, see Appendix~\ref{app:two_qubits}. 

\medskip

\begin{figure}[!t]
    \centering
    \includegraphics[trim=0.2cm 0.2cm 0.1cm 0.2cm, clip, width=0.99\columnwidth]{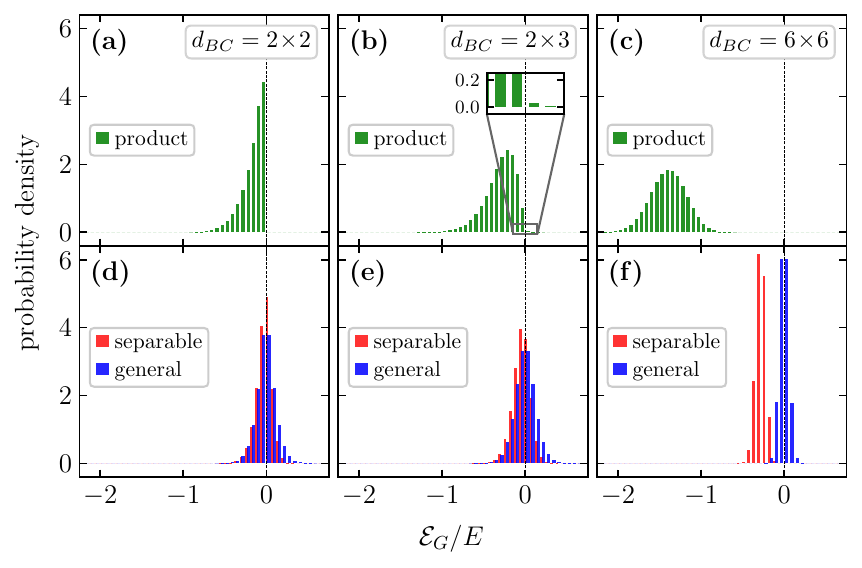}
    \caption{Distribution of ergotropy gain $\erg_\G/E$ for randomly sampled initial states, Hamiltonians $H_{BC}$, and energy-conserving unitaries $U_{BC}$. In panels \textbf{(a)}, \textbf{(b)}, and \textbf{(c)}, $\rho_{BC}$ is a product of random $\rho_B$ and $\rho_C$. The states are drawn from the Hilbert--Schmidt measure, $H_{BC}$ are sampled from the Gaussian Unitary Ensemble, and $U_{BC}$ are sampled from the Haar measure. We see that the ensemble average of $\erg_\G/E$ decreases as the Hilbert space dimension $d_{BC}$ grows.
    Each of the panels \textbf{(d)}, \textbf{(e)}, and \textbf{(f)} shows two histograms of $\erg_\G/E$: one for separable and one for general states of $BC$ ($H_{BC}$ and $U_{BC}$ sampled as above).
    Each histogram depicts $10^6$ points.
    We see that, in all dimensions, the distance between product and general states is significantly larger than the distance between separable and general states.}
    \label{fig:rand_ProdSepGen}
\end{figure}

As noted above, one cannot extend Lemma~\ref{thm:2qbt} to higher Hilbert-space dimensions (see Appendix~\ref{app:two_qubits} for an explicit example). However, as our extensive numerics show, even in higher dimensions correlations play a central role in ergotropy transport. This is vividly seen in Fig.~\ref{fig:rand_ProdSepGen}. Panels (a), (b), and (c) are histograms of $\erg_\G/E$ calculated for a large sample of randomly sampled product states on randomly generated $H_{BC}$, acted upon by random $U_{BC}$. The quantity $E$ is a characteristic energy scale of the system. For qubits, $E$ is simply the energy gap. In higher dimensions, however, multiple energy scales exist, and no single scale can be picked out. This makes defining dimensionless energetic quantities more involved, and $\erg_\G/E$ is a shorthand notation for the dimensionless ergotropy gain (see Appendix~\ref{app:randham} for details). On panel (a), for $B$ and $C$ being two qubits, we see that $\erg_\G/E$ never becomes positive, in accordance with Lemma~\ref{thm:2qbt}. Panels (b) and (c) show that, for higher-dimensional $B$ and $C$, although $\erg_\G/E$ is sometimes positive, its bulk becomes more and more negative as $d_{BC}$ grows. We numerically analyze this in detail in Appendix~\ref{app:ProdStates_HighDim}. We find that the ensemble average of $\erg_\G/E$ is indeed negative and its magnitude grows polynomially with $d_{BC}$.

This thus demonstrates that lack of correlations in the initial state is unfavorable for ergotropy transport. This motivates two questions: Which type of correlations matter the most? Is there a quantitative link between their strength and ergotropy gain? Figures~\ref{fig:rand_ProdSepGen}(d)--(f) shed light on the first question. Each of these panels contains histograms of two ensembles of initial states: random (general) states of $BC$ and random separable states of $BC$. As above, here $H_{BC}$ and $U_{BC}$ are also sampled randomly. We see that, for low-dimensional $B$ and $C$, in terms of ergotropy gain, separable states perform almost as well as all possible bipartite states, including the entangled ones. In higher dimensions, the distance between the bulks of separable and general states is much smaller than the distance between general and product states. Thus, although entangled states are more advantageous for ergotropy transport, especially in high dimensions, the presence of any type of correlations has a much bigger impact on $\erg_\G/E$ than their type. Note that we plot separable vs. general, and not separable vs. entangled, because efficiently characterizing the set of entangled states is impossible for $d_{BC} > 6$ \cite{Horodecki_2009}. However, it is possible to sample the separable set in arbitrary dimensions. In Appendix~\ref{app:randstate}, based on a pure-state decomposition of separable states from Ref.~\cite{Horodecki_1997}, we construct an efficient algorithm of generating random separable states that reasonably uniformly cover the whole separable set.

\medskip

Now, having determined that the general presence of correlations is the primary factor affecting the ergotropy gain, we are ready to answer the second question above. To do so, we use the standard quantifier of ``general'' (or ``total'') correlations---the quantum mutual information \cite{Nielsen_book_2010}: $I(\rho_{BC}) = S(\rho_B) + S(\rho_C) - S(\rho_{BC})$. Since $\erg_\G$ is given by the change in the ergotropic gap [see Eq.~\eqref{gain_vs_gap}], we will compare $\erg_\G$ with the \textit{change} in the mutual information during transport: $\Delta I = I(\tilde{\rho}_{BC}) - I(\rho_{BC})$. In view of the unitary invariance of the von Neumann entropy, we have
\begin{align} \label{QMIchange}
    \Delta I = S(\tilde{\rho}_B) + S(\tilde{\rho}_C) - S(\rho_B) - S(\rho_C).
\end{align}

\begin{figure}[!t]
    \centering
    \includegraphics[trim=0.2cm 0.3cm 0.1cm 0.2cm, clip, width=0.99\columnwidth]{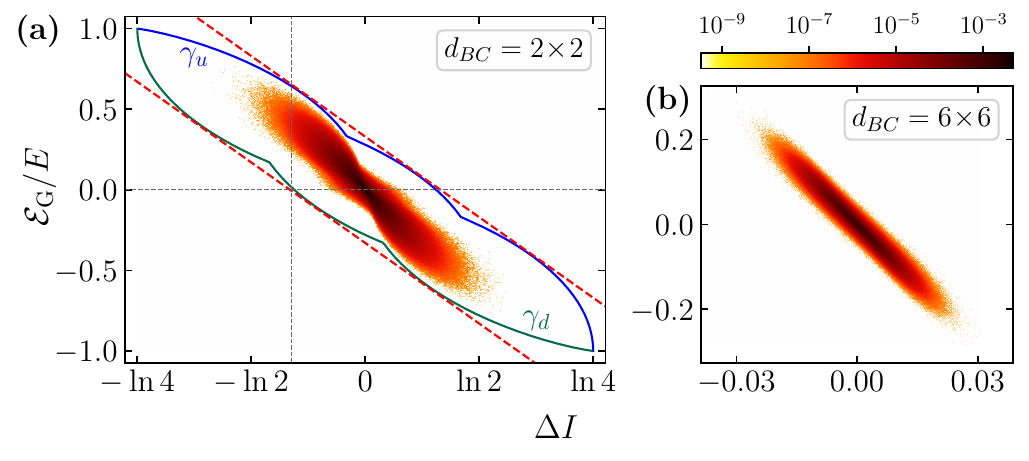}
    \caption{2D histogram of $\mathcal{E}_\G/E$ vs $\Delta I$ for general $\rho_{BC}$, $H_{BC}$, and $U_{BC}$ randomly generated as in Fig.~\ref{fig:rand_ProdSepGen}.
    Panel \textbf{(a)} is for ergotropy transport between two qubits. Here the points distribute in a ``propeller shape''. The blue and green curves $\gamma_u$ and $\gamma_d$ provide tight upper and lower analytical bounds on the distribution. The cumbersome closed-form expressions for them are derived in Appendix~\ref{app:propeller_bounds}. The red dashed lines, which are the tightest linear bounds, show that $\erg_\G/E$ is guaranteed to be positive whenever $\Delta I < - 2 \ln (5 / 4)$; the dotted grey lines cross at this turning point.
    Panel \textbf{(b)} illustrates what happens in higher dimensions, showcasing the $d_B = d_C = 6$ case. Compared to panel (a), the distribution is more concentrated around $0$. This reflects the fact that both $\mathcal{E}_\G/E$ and $\Delta I$ asymptotically concentrate around their ensemble averages.}
    \label{fig:Erg-MI_propeller}
\end{figure}

To reveal how the two quantities are related, we, as above, randomly sample a large ensemble of $\rho_{BC}$, $H_{BC}$, and $U_{BC}$, and plot $\erg_\G/E$ against $\Delta I$. This is done in Fig.~\ref{fig:Erg-MI_propeller}, where $2\times 2$ [panel (a)] and $6\times 6$ [panel (b)] dimensional cases are depicted. For the two qubits, it reveals a propeller-like shape for the distribution. Its width is nonzero, therefore there is no deterministic relation between $\Delta I$ and $\erg_\G/E$.

In higher dimensions, the ranges of $\Delta I$ and $\erg_\G/E$ shrink, as is evidenced by Fig.~\ref{fig:Erg-MI_propeller}(b). In Appendix~\ref{app:levy}, we show that this phenomenon is general, by proving Levy-type measure-concentration bounds \cite{Milman_book_2001, Watrous_book_2018} for both $\erg_G/E$ and $\Delta I$ considered as functions of the initial state $\rho_{BC}$. There, and throughout the Letter, we sample the states from the Hilbert--Schmidt measure \cite{Bengtsson_book_2006, Zyczkowski_2001}. Our proof is based on the fact that both the ergotropy and von Neumann entropy, as functions of the state, are Lipschitz continuous with respect to the Bures metric \cite{Sekatski_2021, Hovhannisyan_2024b}. Our concentration bounds establish that the widths of the distributions of $\Delta I$ and $\erg_\G/E$
decrease polynomially with $d_{B}$ and $d_{C}$. Furthermore, as we numerically demonstrate in Appendix~\ref{app:HighDim_Num}, the relation between $\Delta I$ and $\erg_\G/E$ becomes increasingly ``deterministic''---for a given $\Delta I$, the dispersion of $\erg_\G/E$ decreases with increasing dimensions---even when adjusted for the shrinkage due to concentration.

The concentration of $\Delta I$ and $\erg_\G$ means that, on the one hand, tailoring initial states and transfer unitaries such that ergotropy gain is significant becomes increasingly more complex as the dimensions grow higher. On the other hand, generic transport is robust in that both losses (and gains) are severely restricted.

\medskip

\textit{Multiple transports and channel reusability.}---As we saw above, correlations generically endow the initial state of $BC$ with a positive ergotropic gap $\delta^\erg_{BC}$, allowing one to avoid losses during transport. As Eq.~\eqref{gain_vs_gap} shows, each time $\erg_\G > 0$, this resource gets used up by that exact amount. While for a single transport having the largest $\erg_\G$ is best, the situation changes when one wishes to perform multiple transports.

The latter is arguably the most relevant practical scenario, where our setup is a wire through which third parties send ergotropy to fourth parties, aiming for the highest number of lossless transports. Thus, having $\erg_\G$ \textit{positive but small} would be desirable, so that the initial supply of $\delta^\erg_{BC}$ is consumed slowly. A key quantity here is the total amount of ergotropy, $\ergt^+$, that one can losslessly transport through the channel in multiple iterations. Labeling transport iterations by the index $\iota$,
\begin{align} \label{erg_ext_total}
    \ergt^+ := \sum_{\iota; \; \erg_\G^{(\iota)} \geq 0} \big[ \erg\big(\tilde{\rho}_C^{(\iota)}\big) - \erg\big(\rho_C^{(\iota)}\big) \big].
\end{align}

A paradigmatic protocol realizing this scenario is as follows. The system starts locally empty: both $\rho^{(0)}_B$ and $\rho^{(0)}_C$ are passive ($\rho^{(0)}_{BC}$ may be correlated). Then, a ``charging'' unitary $U_\cha$ acts locally on $B$, injecting some ergotropy into it. Next, an energy-conserving $U_{BC}$ is performed on the total system, transporting ergotropy from $B$ to $C$. Finally, a ``draining'' unitary $U_\dra$ acts locally on $C$, extracting (part of) its ergotropy. This comprises a full ``transport cycle'' of the channel, and is realized by the unitary operation
\begin{align}
    U_\cyc = (\id_B \otimes U_\dra) \, U_{BC} \, (U_\cha \otimes \id_C).
\end{align}
As noted above, the ergotropic gap $\delta^{(0)} := \delta^\erg_{BC}\big(\rho^{(0)}_{BC}\big)$ of the initial ``locally empty'' state is \textit{the} resource that will fuel the gainfulness of the transport and be consumed over time. Since the ergotropic gap depends only on the local and global spectra of the state [see Eq.~\eqref{loc_inv_erg_gap}], the consumption of $\delta^{(0)}$ is only determined by $U_{BC}$.

\medskip

For a transparent exemplification of this situation, we again turn to qubits and take $\rho_{BC}^{(0)} = \ket{0}_B\!\bra{0} \otimes \ket{0}_C\!\bra{0}$, $U_\cha = U_\dra = \sigma_X$, and $U_{BC} = \mathrm{SWAP}$, where $\sigma_X$ is the Pauli-$X$ matrix. In this idealized scenario, the state returns to $\rho_{BC}^{(0)}$ after each transport cycle, and although $\erg_\G = 0$, an $E$ amount of ergotropy is injected into $B$ and the same $E$ amount of ergotropy is extracted from $C$ each time. Thus, this noiseless, idealized channel can be used infinitely many times to losslessly transport an infinite amount of ergotropy.

We caution, however, that this behavior is not stable: a small deviation of $U_{BC}$ from a swap will make the transport lossy. Indeed, let us introduce a small $\varepsilon$ error in the transport unitary:
\begin{align}
    U^{(\varepsilon)}_{BC} = \mathrm{SWAP} + 2 \nu^{(\varepsilon)}_{BC} \sin(\varepsilon/2).
\end{align}
Here $\nu^{(\varepsilon)}_{BC} = \bar{\sigma}_Z \cos\frac{\varepsilon}{2} - \bar{\sigma}_X \sin\frac{\varepsilon}{2}$, where $\bar{\sigma}_X$ and $\bar{\sigma}_Z$ are the Pauli $X$ and $Z$ matrices in the subspace spanned by $\ket{0}_B \otimes \ket{1}_C$ and $\ket{1}_B \otimes \ket{0}_C$. Keeping $U_\cha = U_\dra = \sigma_X$, it is easy to calculate that, upon iteration $\iota$ of the transport cycle, the ergotropy gain $\erg_\G^{(\iota)} = - 2 E \sin(2 \iota \varepsilon - \varepsilon) \sin \varepsilon$, which is negative for $1 \leq \iota \leq \pi/(2 \varepsilon) + 1/2$.

To combat this lossiness, let us take an initial state with a positive ergotropic gap. For example, consider $\rho_{BC}^{(0)} = |\phi^{(0)}\rangle \langle\phi^{(0)}|$ with $|\phi^{(0)}\rangle = \cos \kappa \, \ket{0}_B \otimes \ket{0}_C + \sin \kappa \, \ket{1}_B \otimes \ket{1}_C$, where $0 < \kappa \leq \pi/4$ controls the initial ergotropic gap: $\delta^{(0)} = 2 E \sin^2 \kappa > 0$.
In Appendix~\ref{app:reuse}, we show that, in this case,
\begin{align} \label{ergGiota}
    \erg_\G^{(\iota)} = 2 E \sin(2 \kappa + \varepsilon - 2 \iota\varepsilon) \sin \varepsilon.
\end{align}
This quantity is positive for the first $\left\lfloor \tfrac{\kappa}{\varepsilon} + \tfrac{1}{2}\right\rfloor$ iterations ($\lfloor x \rfloor$ is the floor function). For example, for $\varepsilon = 0.03$ and $\kappa = \pi/8$, the transport is gainful during the first $13$ iterations. As a result, $\ergt^+ \approx 11.84 E$ is collected from $C$.
Thus, a single slightly correlated state can be used to losslessly transport an order of magnitude more work ($\ergt^+$) than it can store ($E$).

We emphasize that, although the transport is executed multiple times, the correlated state $\rho_{BC}^{(0)}$ is prepared only once. This is important because correlations rarely come for free---creating them costs energy \cite{Huber_2015, Bera_2017, Vitagliano_2018, Bakhshinezhad_2019}. For the example above, the work cost of creating $|\phi^{(0)} \rangle$ from the uncorrelated state $\ket{0}_B \otimes \ket{0}_C$ by unitary operations is $2 E \sin^2\kappa = \delta^{(0)} \approx 0.29 E$. And it is remarkable that this small investment unlocks $\approx 40.82$ times as much losslessly transportable ergotropy (recall that $\erg^+_{\mathrm{tot}} \approx 11.84 E$).

Lastly, we note that our multiple transfer setup is fundamentally different from that considered in Ref.~\cite{Lobejko_2022}. Using our nomenclature, only uncorrelated, incoherent (diagonal) initial states are considered in Ref.~\cite{Lobejko_2022}, and a fresh battery is used upon each iteration.

\medskip

\textit{Summary and outlook.}---We introduced a paradigmatic setup of a quantum battery autonomously transporting ergotropy to a quantum device (consumption unit). We established that initial battery--device correlations can significantly enhance the transport. Despite the channel being energy conserving, the device may get more ergotropy than the battery loses. Such gainful transport is counterintuitive, as the natural expectation is that, when transported, a useful resource should suffer dissipation.

We proved that, between two qubits, gainful transport is possible \textit{only} if they are initially correlated. Although for higher-dimensional systems the correlation--ergotropy-gain relation becomes fuzzier, we showed that there, too, correlations maintain a central role.
We also established that $\Delta I$ and $\erg_\G$ concentrate with increasing dimensions, meaning that ergotropy transport between large systems is generically robust---while large gains are not to be expected, large losses are also unlikely.

Finally, we showed that a one-time creation of correlations between $B$ and $C$ can have a long-lasting loss prevention effect when, subsequently, multiple iterations of an imperfect channel transport ergotropy through $BC$.
Even a modest amount of correlations between $B$ and $C$ can enable lossless transport of an order of magnitude more work than
the energetic resources needed to create the initial correlated state.

The framework developed here may serve as a stepping stone for further research into ergotropy transport under realistic conditions. A significant step would be accounting for environmental noise during transport, e.g., by considering non-unitary energy-conserving channels \cite{Chiribella_2017}.

\medskip

We thank Carsten Henkel and Marcin \L{}obejko for useful discussions.
K.V.H. and J.A. are grateful for support from the University of Potsdam.
J.A. gratefully acknowledges funding from the Deutsche Forschungsgemeinschaft (DFG, German Research Foundation) under Grants No. 384846402 and No. 513075417 and from the Engineering and Physical Sciences Research Council (EPSRC) (Grant No. EP/R045577/1) and thanks the Royal Society for support.

\medskip

The code used to produce the data shown in Figs.~\ref{fig:rand_ProdSepGen}--\ref{fig:SM-propeller} is available upon reasonable request to K.V.H., \href{mailto:karen.hovhannisyan@uni-potsdam.de}{karen.hovhannisyan@uni-potsdam.de}.



\bibliography{literature.bib}


\appendix

\section{Proof of Lemma~\ref{thm:2qbt}}
\label{app:two_qubits}

Before proceeding to the core proof of Lemma~\ref{thm:2qbt}, let us first show that the ergotropic gap $\delta^\erg_{BC}$ depends only on the spectra of the global and local states whenever $B$ and $C$ do not interact [i.e., the total Hamiltonian is of the form given by Eq.~\eqref{totham}]. This important property of the ergotropic gap will become useful in what follows. To prove it, let us first invoke the well-known fact that the minimum in Eq.~\eqref{ergot_def_1} is delivered by the unitary that rotates $\rho$ to the so-called passive state $\rho^\downarrow$ \cite{Pusz_1978, Lenard_1978, Allahverdyan_2004}, and thus
\begin{align} \label{ergot_def_2}
	\erg(\rho) = \tr[(\rho - \rho^\downarrow) H].
\end{align}
The passive state $\rho^\downarrow = \sum_k \lambda_k \ket{E_k}\bra{E_k}$, where $\ket{E_k}$ are the eigenvectors of $H$ for which the corresponding eigenvalues are ordered increasingly ($E_0 \leq \cdots \leq E_{d-1}$) and $\lambda_0 \geq \cdots \geq \lambda_{d-1} \geq 0$ are the eigenvalues of $\rho$ in decreasing order. Note that, whenever $\rho$ or $H$ have degenerate spectra, $\rho^\downarrow$ will not be unique.
 
Now, using Eq.~\eqref{ergot_def_2}, it is straightforward to show that
\begin{align} \label{loc_inv_erg_gap}
    \delta^\erg_{BC} = \tr \big[ \big( \rho_B^\downarrow \otimes \rho_C^\downarrow - \rho_{BC}^\downarrow \big)H_{BC} \big],
\end{align}
where $\rho_{BC}^\downarrow$ is passive with respect to $H_{BC}$ and $\rho_X^\downarrow$ is passive with respect to $H_X$ ($X = B, \, C$).

\medskip

To prove Lemma~\ref{thm:2qbt}, let us without loss of generality set the energies of the ground states of $B$ and $C$, $\ket{0}_B$ and $\ket{0}_C$, to zero. With this convention, the Hamiltonians write as $H_B = E_B \ket{1}_B \!\bra{1}$ and $H_C = E_C \ket{1}_C \! \bra{1}$.

There are two qualitatively different cases with regard to the structure of the energy-conserving unitary: $E_B \neq E_C$ and $E_B = E_C := E$. We treat them separately.

\textbf{Case 1: $E_B \neq E_C$}. In this case, $H_{BC} = H_B \otimes \id_C + \id_B \otimes H_C$ is nondegenerate, and therefore only operators that are diagonal in the energy eigenbasis can satisfy Eq.~\eqref{erg_con_uni}. Thus, in the energy eigenbasis, energy-conserving unitaries are necessarily of the form
\begin{align}
    U_{BC} = \diag(e^{i \phi_0}, e^{i \phi_1}, e^{i \phi_2}, e^{i \phi_3}).
\end{align}
It is now a simple exercise to check that the elements of the post-transfer state of $B$, $\tilde{\rho}_B = \tr_C \big[ U_{BC} \rho_B \otimes \rho_C U_{BC}^\dagger \big]$, are
\begin{align*}
    (\tilde{\rho}_B)_{00} &= (\rho_B)_{00},
    \\
    (\tilde{\rho}_B)_{01} &= (\rho_B)_{01} \, \big[ (\rho_C)_{00} e^{i (\phi_0 - \phi_2)} + (\rho_C)_{11} e^{i (\phi_1 - \phi_3)} \big].
\end{align*}
Due to the triangle inequality,
\begin{align*}
    \big\vert (\rho_C)_{00} e^{i (\phi_0 - \phi_2)} + (\rho_C)_{11} e^{i (\phi_1 - \phi_3)} \big\vert \leq (\rho_C)_{00} + (\rho_C)_{11} = 1.
\end{align*}
Therefore,
\begin{align} \label{dephB}
    \big\vert (\tilde{\rho}_B)_{01} \big\vert \leq \big\vert (\rho_B)_{01} \big\vert.
\end{align}
Similarly, for $\tilde{\rho}_C = \tr_B \big[ U_{BC} \rho_B \otimes \rho_C U_{BC}^\dagger \big]$, we have
\begin{align*}
    (\tilde{\rho}_C)_{00} &= (\rho_C)_{00},
    \\
    (\tilde{\rho}_C)_{01} &= (\rho_C)_{01} \, \big[ (\rho_B)_{00} e^{i (\phi_0 - \phi_1)} + (\rho_B)_{11} e^{i (\phi_2 - \phi_3)} \big],
\end{align*}
and
\begin{align} \label{dephC}
    \big\vert (\tilde{\rho}_C)_{01} \big\vert \leq \big\vert (\rho_C)_{01} \big\vert.
\end{align}

Now, let us recall that the ergotropy of a qubit with the Hamiltonian $E \ket{1}\bra{1}$ is
\begin{align}
    \erg(\rho) = E \Big( \frac{1}{2} - \rho_{00} + \sqrt{\frac{1}{4} - \rho_{00} \rho_{11} + |\rho_{01}|^2} \Big).
\end{align}
Thus, $\erg$ is a monotonically increasing function of $|\rho_{01}|$. Therefore, in view of Eqs.~\eqref{dephB} and~\eqref{dephC},
\begin{align}
    \erg(\tilde{\rho}_B) \leq \erg(\rho_B) \quad \mathrm{and} \quad \erg(\tilde{\rho}_C) \leq \erg(\rho_C),
\end{align}
which in turn means that, according to its definition in Eq.~\eqref{erggain}, the ergotropy gain $\erg_\G \leq 0$.

\textbf{Case 2: $E_B = E_C := E$}. Here the spectrum of $H_{BC}$ is degenerate---$\spec(H_{BC}) = \{0, E, E, 2E\}$---and hence there exist nontrivial energy-conserving unitaries. However, in this case, $\rho_B^\downarrow \otimes \rho_C^\downarrow$ is passive with respect to $H_{BC}$, and therefore $\delta^\erg_{BC} = 0$ due to Eq.~\eqref{loc_inv_erg_gap}. Indeed, denoting the eigenvalues of $\rho_B$ and $\rho_C$ by, respectively, $\lambda^B_0 \geq \lambda^B_1$ and $\lambda^C_0 \geq \lambda^C_1$, we have that
\begin{align*}
    \rho_B^\downarrow \otimes \rho_C^\downarrow = \diag\big(\lambda^B_0 \lambda^C_0, \lambda^B_0 \lambda^C_1, \lambda^B_1 \lambda^C_0, \lambda^B_1 \lambda^C_1\big)
\end{align*}
in the energy eigenbasis. And since $\lambda^B_0 \lambda^C_0 \geq \lambda^B_0 \lambda^C_1 \geq \lambda^B_1 \lambda^C_1$ and $\lambda^B_0 \lambda^C_0 \geq \lambda^B_1 \lambda^C_0 \geq \lambda^B_1 \lambda^C_1$, this state is passive with respect to $H_{BC} = \diag(0, E, E, 2E)$. Thus, due to Lemma~\ref{thm:lossy}, ergotropy transport by energy-conserving unitaries acting on product states cannot be gainful. This concludes the proof of Lemma~\ref{thm:2qbt}.

\medskip

Two remarks are due here. First, while Lemma~\ref{thm:2qbt} proves that correlations are necessary for $\erg_\G > 0$, their presence is not sufficient for achieving positive ergotropy gain. For example, when $E_B = E_C := E$, the state
\begin{align*}
	\frac{1}{2} \ket{0}_B\!\bra{0} \otimes \ket{0}_C\!\bra{0} + \frac{1}{2} \, |\Psi^+\rangle \langle \Psi^+|,
\end{align*}
where $\ket{\Psi^+} = (\ket{0}_B \otimes \ket{1}_C + \ket{1}_B \otimes \ket{0}_C) / \sqrt{2}$, is entangled but has zero ergotropic gap. Note that, since entangled states always have nonzero quantum discord \cite{Dakic_2010}, this example may appear to contradict the claim of Ref.~\cite{Mukherjee_2016} that states with nonzero discord have positive ergotropic gap. However, there is no contradiction: in our case, the qubits have identical energy spectra, whereas the result of Ref.~\cite{Mukherjee_2016} applies only when the spectra are different.

Second, due to the activation phenomenon mentioned in the main text, we know that Lemma~\ref{thm:2qbt} cannot be directly generalized to higher dimensions. The minimal counterexample is found in $2 \times 3$ dimensions. Take $B$ and $C$ with Hamiltonians $H_B = E \ket{1}_B\!\bra{1}$ and $H_C = E \ket{1}_C\!\bra{1} + E \ket{2}_C\!\bra{2}$. Then, $\rho_B \otimes \rho_C$, with $\rho_B = \frac{1}{2} \id_B$ and $\rho_C = \tfrac{1}{2} \ket{0}_C\!\bra{0} + \tfrac{1}{2} \ket{1}_C\!\bra{1}$, is not passive despite both $\rho_B$ and $\rho_C$ being passive. In the extreme case when $C$ is a large thermal bath, a natural family of such counterexamples was identified in Ref.~\cite{Biswas_2022}.

\medskip

Finally, in Appendix~\ref{app:QuanMargProb}, we provide an alternative proof of Lemma~\ref{thm:2qbt} by making a connection to the so-called quantum marginal problem \cite{Bravyi_2003, Klyachko_2006, Christandl_2005} and using its solution for two qubits \cite{Bravyi_2003}. In fact, we go even deeper and show that this connection can be useful in both ways, by deriving a family of nontrivial inequalities for quantum marginals \textit{in arbitrary dimensions} from the simple fact that $\delta^\erg_{BC} \geq 0$.

\section{Details of our numerical simulations}
\label{app:numerics}

All thermodynamic and information-theoretic quantities that we calculate in this paper are fully determined by the initial state $\rho_{BC}$, the local Hamiltonians $H_B$ and $H_C$, and the energy-conserving transport unitary $U_{BC}$. Here we give a detailed exposition of how we sample random states, Hamiltonians, and energy-conserving unitaries in our numerical simulations. Out of many possible ways to randomly sample these quantities, we aimed at those that would cover the relevant sets fully, be physically meaningful, and mathematically well-defined.

The high-level procedure we follow is: (i) Generate a random $\rho_{BC}$ (sometimes with a condition on the correlations). (ii) Independently from $\rho_{BC}$, generate a random $H_{BC}$ of the form in Eq.~\eqref{totham}, with the condition that its spectrum has nontrivial degeneracies. (iii) Generate a random energy-conserving transport unitary $U_{BC}$ that operates in the degenerate eigensubspaces of $H_{BC}$. Obviously, the sampling in step (iii) strongly depends on the particular outcome of the sampling in step (ii).

\subsection{Sampling initial states}
\label{app:randstate}

There are two important desiderata that a sensible probability measure on the space of density matrices should satisfy. First of all, in order to be analytically and numerically tractable, we require it to be induced by a Riemannian metric so that the volume element is constructed intuitively and can be written explicitly. Second, for such a metric---let us call it $\D$---we require unitary invariance: $\D(U\rho U^\dagger, U \sigma U^\dagger) = \D(\rho, \sigma)$ for arbitrary states $\rho$ and $\sigma$ and unitary $U$. These requirements ensure that $\D$ is Fubini--Study--adjusted \cite{Bengtsson_book_2006}. Namely, on pure states, $\D$ coincides with the Fubini--Study metric, since the latter is the only (up to a constant factor) unitarily invariant Riemannian metric on pure states \cite{Bengtsson_book_2006}. As a result of unitary invariance, the measure induced by the Fubini--Study metric is uniform on pure states \footnote{Any pure state in a $d$-dimensional Hilbert space with some orthonormal basis $\{ \ket{k}\}_{k=1}^d$ can be decomposed as $\sum_k c_k \ket{k}$, and the normalization condition means $\sum_k |c_k|^2 = 1$. This defines a $(2d-1)$-dimensional sphere, and the Fubini--Study measure on pure states is simply the Haar measure on this sphere.}. Conveniently, random states of the form $U\ket{0}$, where $\ket{0}$ is some arbitrarily chosen pure state and $U$ is sampled from the Haar measure on the unitary group $U(d)$, are distributed according to the Fubini--Study measure \cite{Bengtsson_book_2006}.

Among infinitely many unitarily invariant Riemannian metrics on the manifold of quantum states \cite{Bengtsson_book_2006}, the Hilbert--Schmidt and Bures metrics stand out. The former, defined as
\begin{align}
    \DHS(\rho, \sigma) = \sqrt{\tr[(\rho - \sigma)^2]},
\end{align}
is the generalization of the standard Euclidean metric for vectors to matrices, and it is widely used in quantum information \cite{Nielsen_book_2010, Bengtsson_book_2006, Watrous_book_2018} and computing \cite{Travnicek_2019, Liu_2020, Cerezo_2021} since it is intuitive and efficiently computable. Although it does not have a direct operational meaning, it has been used to quantify coherence and entanglement \cite{Dodonov_2000, Pandya_2020}. Moreover, it is contractive (or ``monotone'') under the action of unital channels \cite{Perez-Garcia_2006} (but not under more general completely positive trace-preserving maps \cite{Ozawa_2000}).

Crucially, the measure induced by the Hilbert--Schmidt metric is also natural and intuitive. Indeed, take two copies of the system ($B$ and $B'$) and draw random pure states $\ket{\Psi}$ in the joint $d^2$-dimensional Hilbert space from the Fubini--Study measure. Then, the density matrices $\tr_{B'} \ket{\Psi}\!\bra{\Psi}$ will be distributed according to the Hilbert--Schmidt measure in the Hilbert space of $B$ \cite{Zyczkowski_2001}. Numerically, a more efficient way of sampling random density matrices from the Hilbert--Schmidt measure is to sample a $d\times d$ random Ginibre matrix $G$ and compute
\begin{align} \label{HS_rs}
    \rho = \frac{G G^\dagger}{\tr (G G^\dagger)}.
\end{align}
As was proven in Ref.~\cite{Zyczkowski_2001}, the states $\rho$ are indeed distributed according to the Hilbert--Schmidt measure. We remind the reader that the elements of random Ginibre matrices are independent, identically identically distributed Gaussian random variables with mean $=0$ and variance $=1$ \cite{Mezzadri_2007}.

The other prominent metric---the Bures metric---defined as \cite{Bures_1969} $\DBu(\rho, \sigma) = \sqrt{2 - 2 \sqrt{F(\rho, \sigma)}}$, where $F(\rho, \sigma) = \big(\tr\sqrt{\rho^{1/2} \sigma \rho^{1/2}}\big)^2$ is the Uhlmann fidelity \cite{Uhlmann_1976}, has additional desirable properties that the Hilbert--Schmidt metric does not have. Namely, $\DBu$ is contractive under the action of completely positive trace-preserving maps \cite{Petz_1996, Bengtsson_book_2006}
and Fisher-adjusted \cite{Braunstein_1994, Bengtsson_book_2006} (i.e., for diagonal states, it coincides with the Fisher information metric \cite{Amari-Nagaoka_book_2000}). In fact, $\DBu$ is the only unitarily invariant Riemannian metric that has these properties \cite{Petz_1996a, Sommers_2003, Bengtsson_book_2006}. Due to these unique properties, the measure induced by the Bures metric is considered \cite{Slater_1996, Slater_1998, Sommers_2003} to be the quantum analog of the Jeffreys prior \cite{Amari-Nagaoka_book_2000}. The latter is the least informative distribution in classical statistics \cite{Amari-Nagaoka_book_2000} under certain generic conditions. Thus, the Bures measure is ''the most random'' measure, and it is a natural choice for revealing \textit{generic} properties of states.

Despite these desirable properties, the curvature of the manifold of states in Bures metric diverges as the points of rank change \cite{Dittmann_1995, Bengtsson_book_2006}. Moreover, the Bures measure is skewed towards pure states as compared to the Hilbert--Schmidt measure \cite{Hall_1998, Zyczkowski_2001, Bengtsson_book_2006}. Thus, while the Bures measure is more random than the Hilbert--Schmidt measure, the latter is smooth and overall more ``flat'' in terms of purity. At the same time, the two measures are not too different in that they produce similar results in several contexts \cite{Sarkar_2019, Slater_2019, Wei_2022}.

\medskip

In our analysis, we require the measure from which states are sampled to be as random as possible, while guaranteeing a fair representation for density matrices of all ranks. Given all of the above, we believe the Hilbert--Schmidt measure strikes a good balance in that it (i) is natural in the sense that it is merely a reduction of the Fubini--Study measure, (ii) is close to the maximally random (Bures) measure, (iii) produces a flatter distribution with respect to purity than the Bures measure.

Therefore, in our numerical simulations, whenever we need to generate a random state of $BC$ without any restrictions, we simply sample a $d_{BC} \times d_{BC}$ random density matrix from the Hilbert--Schmidt measure using Eq.~\eqref{HS_rs}. Whenever we need to generate an uncorrelated state of the form $\rho_B \otimes \rho_C$, we sample $\rho_B$ and $\rho_C$ independently, each from the Hilbert--Schmidt measure.

The situation becomes more subtle when we need to generate random separable states of $BC$. Whenever $d_{BC} \leq 6$, doing so is relatively straightforward because it is easy to check whether a state is separable using the Peres criterion, i.e., partially transposing the state and checking if it is positive-semidefinite \cite{Horodecki_2009}. Thus, in order to generate a random separable state in $d_{BC} \leq 6$ dimensions, we can simply keep sampling random $\rho_{BC}$'s until we hit a separable state. Let us call this method Peres-filtered Hilbert--Schmidt (PFHS) sampling. This method is used to sample separable states in Figs.~\ref{fig:rand_ProdSepGen}(d) and~(e).

\begin{figure}[!t]
    \centering
    \includegraphics[width=\columnwidth]{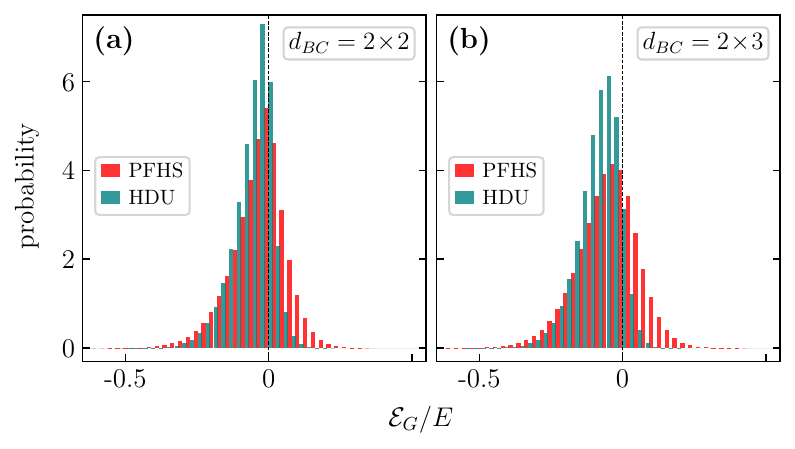}
    \caption{Comparing separable-state-sampling algorithms. Depicted are histograms of $\erg_\G$ for randomly sampled separable initial states $\rho_{BC}$, Hamiltonians $H_{BC}$, and energy-conserving unitaries $U_{BC}$ (similarly to Fig.~\ref{fig:rand_ProdSepGen}). Each histogram is made up of $10^6$ points. Panel \textbf{(a)} is for $d_{BC} = 2 \times 2$. The red histogram results from Peres-filtered Hilbert--Schmidt (PFHS) sampling of $\rho_{BC}$, whereas the green the histogram results from the Horodecki-decomposed uniform (HDU) sampling algorithm. Panel (b) does the same as (a), but for $d_{BC} = 2 \times 3$. In both configurations, the distributions are fairly similar, although the one based on Eq.~\eqref{rand_sep} is slightly more concentrated. The color red for the PFHS sampling is chosen for consistency with Fig.~\ref{fig:rand_ProdSepGen}, where, for $2\times 2$ and $2\times 3$ dimensions, separable states are sampled from PFHS.}
    \label{fig:Werner-vs-Peres}
\end{figure}

However, when $d_{BC} > 6$, checking whether a state is separable is unfeasible \cite{Horodecki_2009}, so the mentioned technique becomes inapplicable. To go beyond this impasse, we use the fact, proven in Ref.~\cite{Horodecki_1997}, that any separable state of $BC$ can be represented as a convex combination of at most $d_{BC}^2$ uncorrelated pure states. Thus, given that sampling from the Fubini--Study measure is the most sensible way of generating pure states, and it is what the Hilbert--Schmidt measure reduces to on pure states, we can generate random separable states as
\begin{align} \label{rand_sep}
    \sum_{a=1}^{d_{BC}^2} p_a \, U_a^{(B)}\ket{0}_B \! \bra{0} U_a^{(B) \, \dagger} \otimes U_a^{(C)}\ket{0}_C \! \bra{0} U_a^{(C) \, \dagger}.
\end{align}
Here $\ket{0}_B$ and $\ket{0}_C$ are some arbitrarily chosen pure states of, respectively, $B$ and $C$. Each random unitary $U_a^{(B)}$ is sampled from the Haar measure on the unitary group $U(d_B)$, and analogously for $U_a^{(C)}$. Haar-distributed unitary matrices can be generated by generating a full-rank Ginibre matrix and performing the Gram--Schmidt orthonormalization procedure to its rows (or columns) \cite{Mezzadri_2007}.

The coefficients $p_a \geq 0$ in Eq.~\eqref{rand_sep} form a $(d_{BC}^2-1)$-dimensional simplex, $\sum_{a=1}^{d_{BC}^2} p_a = 1$, and they are also sampled randomly. Out of several intuitive ways of randomly sampling points from that simplex \cite{Willms_2021}, we choose the uniform one. It is obtained by sampling $d_{BC}^2 - 1$ points $\kappa_a$ from the uniform measure on $[0, 1]$, ordering them so that $0 \leq \tilde{\kappa}_1 \leq \cdots \leq \tilde{\kappa}_{d_{BC}^2-1} \leq 1$, and taking $p_a = \tilde{\kappa}_a - \tilde{\kappa}_{a-1}$, with $\tilde{\kappa}_0 = 0$ and $\tilde{\kappa}_{d_{BC}^2} = 1$ \cite{Devroye_book_1986, Willms_2021}.

Now, since thus generated $p_a$'s cover the whole simplex and $U_a^{(B)}\ket{0}_{B}$'s and $U_a^{(C)}\ket{0}_{C}$'s cover all pure states of, respectively, $B$ and $C$, the random separable states in Eq.~\eqref{rand_sep} cover all separable states of $BC$. Although we do not have a proof that this sampling is uniform on the manifold of separable states, it presumably achieves a reasonable degree of uniformity due to the fact that the distributions of $p_a$'s, $U_a^{(B)}\ket{0}_{B}$'s, $U_a^{(C)}\ket{0}_{C}$'s are as uniform as possible. We call this algorithm Horodecki-decomposed uniform (HDU) sampling. To benchmark it, we compare it with PFHS
sampling in dimensions where the latter works reliably (i.e., $d_{BC} = 4$ and $d_{BC} = 6$). This is demonstrated in Fig.~\ref{fig:Werner-vs-Peres}, where the histograms of the distributions of the ergotropy gain for these two separable-state-sampling algorithms are depicted. This plot shows that the difference between the two samplings is not big, meaning that HDU sampling
generates sensible results. That said, as we can see in Fig.~\ref{fig:Werner-vs-Peres}, it is more concentrated around the average, implying that, with respect to $\erg_\G$, it is slightly ``less uniform'' compared to PFHS
sampling.

\subsection{Sampling Hamiltonians}
\label{app:randham}

As with states, we want to sample Hamiltonians as generically as possible, while maintaining a fair representation for fringe cases. Most important fringe cases for Hamiltonians are those with degenerate spectra. Our strategy here is to start with maximal randomness and then gradually impose pertinent limitations.

Since any Hermitian operator can be a Hamiltonian, our starting point (before imposing limitations specific to our problem) is sampling Hermitian matrices as randomly as possible. To keep the possibilities as wide as possible, it is reasonable to require that all $d^2$ real parameters characterizing $d\times d$ Hermitian matrices to be statistically independent. Moreover, since there is no preferred basis in our problem, the probability measure should be invariant under unitary transformations. Remarkably, these two requirements unambiguously lead us to the Gaussian Unitary Ensemble (GUE) \cite{Edelman_2005}. Moreover, sampling matrices $H$ from the GUE is equivalent \cite{Edelman_2005} to sampling
\begin{align} \label{GUEGinibre}
    H = (G + G^\dagger)/2,
\end{align}
where $G$ is a random Ginibre matrix.

Since in our case the total Hamiltonian is of the form $H_{BC} = H_B \otimes \id_C + \id_B \otimes H_C$ [see Eq.~\eqref{totham}], random Hamiltonian generation will consist of generating $H_B$ and $H_C$ separately. However, since we need $H_{BC}$ to allow for nontrivial transport through energy-conserving unitaries, $H_B$ and $H_C$ must have at least one matching nonzero gap in their spectra. This matching condition means that $H_B$ and $H_C$ \textit{cannot} be statistically independent.

Crucially, when $H_B$ and $H_C$ are sampled from the general GUE [Eq.~\eqref{GUEGinibre}], their eigenvalues can be any real numbers, and the subset of $(H_B, H_C)$ pairs with matching gaps is of measure zero. Therefore, if one simply generates $H_B$ and $H_C$, each according to Eq.~\eqref{GUEGinibre}, then $(H_B, H_C)$ pairs with matching gaps will almost never be encountered. To circumvent this problem, we coarse-grain the spectra of the Hamiltonians as follows. First, we generate a Hermitian matrix $H$ using Eq.~\eqref{GUEGinibre}. Then, we calculate its eigenvalues $E_0 \leq \cdots \leq E_{d-1}$. Next, we choose the ``grain'' size of our coarse-graining, $\epsilon > 0$, and
\begin{align*}
    \text{if} \;\; E_k \in (\epsilon m - \epsilon/2, \,  \epsilon m + \epsilon/2], \;\; m \in \mathbb{Z}, \;\; \text{then} \;\; E_k \to \epsilon m.
\end{align*}
Thus, the coarse-grained spectrum, $\{\widetilde{E}_0 \leq \cdots \leq \widetilde{E}_{d-1}\}$, is given by
\begin{align}
    \widetilde{E}_k = \epsilon \, m_k, \quad \mathrm{with} \quad m_k = \bigg\lceil \frac{E_k}{\epsilon} + \frac{1}{2} \bigg\rceil - 1,
\end{align}
where $\lceil x \rceil$ is the ceiling function. Finally, since the choice of the zero level of energy is arbitrary, we shift the spectrum so that $\widetilde{E}_0 = 0$. All in all, this procedure leaves us with a coarse-grained Hamiltonian $\widetilde{H} = \epsilon \, \diag(0, \, m_1, \, \dots, \, m_{d-1})$, where $0 \leq m_1 \leq \cdots \leq m_{d-1}$ are all integers. The degree of randomness of $\widetilde{H}$ has been limited only by the act of coarse-graining.

Now, coming up with $(\widetilde{H}_B, \widetilde{H}_C)$ pairs with matching coarse-grained gaps is numerically feasible: following the above procedure, we independently sample $\widetilde{H}_B$ and $\widetilde{H}_C$ (the grain size $\epsilon$ being the same in both cases). Then, we check if they have at least one matching nonzero gap. If not, we sample again.

\medskip

So far, we have not mentioned any energy units, and for numerical simulations to be physically meaningful, nothing must depend on the energy unit. Thus, we need a way of obtaining dimensionless energetic quantities (such as Hamiltonians, average energies, ergotropy, etc.) that makes physical sense. Obviously, once one finds a meaningful dimensionless Hamiltonian, then all other energetic quantities become dimensionless automatically.

Finding the ``correct'' dimensionless Hamiltonian is straightforward whenever there is a single energy scale in the system. Then, one simply divides the Hamiltonian (and all energetic quantities along the way) by the energy scale. For example, when the system is a qubit, the energy scale is the spectral gap of its Hamiltonian. Or when the system is an oscillator, the energy scale is its frequency multiplied by the Planck's constant.

However, when the Hilbert space dimension is higher than two and the Hamiltonian is generated randomly, there can be no unique energy scale---some energy gaps in the Hamiltonian will be larger than other gaps by orders of magnitude. This makes the ``procedure'' of obtaining a physically sensible dimensionless Hamiltonian nontrivial and, importantly, non-unique.

Now, the above procedure of sampling random Hamiltonians already generates a dimensionless matrix, and since we want as much randomness as possible, we want to retain the diversity of energy gaps in it. Thus, the only parameter we need to adjust to make it physical meaningful is the spectral radius. To gain an intuition as to how to set a scale for the spectral radius of the Hamiltonian, let us consider three examples where that choice is obvious. \\
\textit{Example 1.}---If $d_B = d_C = 2$, then the only Hamiltonian configuration allowing for nontrivial transport is $H_B = E \ket{1}_B \! \bra{1}$ and $H_C = E \ket{1}_C \! \bra{1}$. Here $E$ sets the energy scale and the dimensionless Hamiltonians are simply $\diag(0, 1)$. \\
\textit{Example 2.}---If $d_B \geq 3$, $H_B$ has equidistant levels, and $C$ is simply a qubit, then the only configuration where transport is possible is $H_B = E \sum_k k \ket{k}_B \! \bra{k}$ and $H_C = E m \ket{1}_C \! \bra{1}$, with $m \leq d_B-1$ a natural number. Obviously, here the dimensionless Hamiltonians are $\diag(0, 1, 2, \dots, d_B-1)$ and $\diag(0, m)$, and $E$ sets the energy scale. \\
\textit{Example 3.}---If $d_B \geq 3$, $H_B = E \sum_{k=1}^{d_B - 1} \ket{k}_B\!\bra{k}$, and $C$ is a qubit, then the only nontrivial possibility is that $H_C = E \ket{1}_C \! \bra{1}$. Here the dimensionless Hamiltonians are $\diag(0, 1, 1, \dots)$ and $\diag(0, 1)$, and, again, $E$ sets the energy scale.

When the gaps of the Hamiltonian are not limited as in the examples above, they can be arbitrarily large. While this is certainly a theoretical possibility, in most real-life many-body systems the energy spectrum is limited, only slowly growing with the number of particles. Motivated by this observation and the three examples above, we will set the scale of the dimensionless Hamiltonian as follows. \\
Step 1: Following the prescription described above, generate a random pair $(\widetilde{H}_B, \widetilde{H}_C)$ with at least one matching nonzero gap. \\
Step 2: Determine whose highest eigenvalue is larger. Suppose it is $C$, i.e., $\widetilde{E}_{d_C-1} > \widetilde{E}_{d_B-1}$. \\
Step 3: Determine $M_C$---the number of distinct eigenvalues of $\widetilde{H}_C$ (i.e., all degenerate eigenvalues count as one). \\
Step 4: Rescale the whole Hamiltonian, $\widetilde{H}_B \otimes \id_C + \id_B \otimes \widetilde{H}_C$, by $(M_C - 1) / \widetilde{E}_{d_C-1}$. This will make sure that the maximal (dimensionless) energy of $C$ is $M_C - 1$, and the rest will transform proportionally. \\
Step 5: If in Step 2 it was a tie, namely, $\widetilde{E}_{d_C-1} = \widetilde{E}_{d_B-1}$, then, for rescaling, choose the system with the largest number of distinct eigenvalues.

It is easy to check that, in the situations described by the examples above, this algorithm outputs the dimensionless Hamiltonians obtained in those examples. We use this Hamiltonian generation algorithm in our numerical simulations. In all our simulations, we set $\epsilon = 0.2$.

Of course, the randomness of sampling with this method is somewhat inhibited by coarse-graining, gap-matching, and rescaling (which are necessary to make the Hamiltonian appropriate for our problem and physically meaningful). Nonetheless, we believe that the vast range of configurations covered by these Hamiltonians is representative, and the sampling is reasonably fair in that it does not have any other biases than these three.

\subsection{Sampling energy-conserving transport unitaries}

Given a $H_{BC}$, sampling a fully random $U_{BC}$ such that $[U_{BC}, H_{BC}] = \nul$ is straightforward. Indeed, due to the commutation condition, $U_{BC}$ can act nontrivially only in the degenerate eigensubspaces of $H_{BC}$. Let $M_{BC}$ be the number of distinct eigenvalues $E^{(BC)}_s$ of $H_{BC}$, and $\{g_s\}_{s=1}^{M_{BC}}$ be the dimensions of the corresponding eigensubspaces. By definition, $\sum_s g_s = d_{BC}$, and due to the gap matching between $B$ and $C$, at least one of $g_s$'s will be $\geq 2$.

Then, for the Hilbert-space partitioning in which
\begin{align*}
    H_{BC} = \bigoplus_{s=1}^{M_{BC}} E^{(BC)}_s \Pi^{(BC)}_s,
\end{align*}
where $\Pi^{(BC)}_s$ are the eigenprojectors of $H_{BC}$, we generate the random energy-conserving unitary as
\begin{align}
    U_{BC} = \bigoplus_{s=1}^{M_{BC}} U_s,
\end{align}
where, for each $s$, $U_s$ is sampled according to the Haar measure from the unitary group $U(g_s)$.

\subsection{Numerics for product initial states at higher dimensions}
\label{app:ProdStates_HighDim}

\begin{figure}[!t]
    \centering
    \includegraphics[trim=0.2cm 0.2cm 0.1cm 0.2cm, clip, width=\columnwidth]{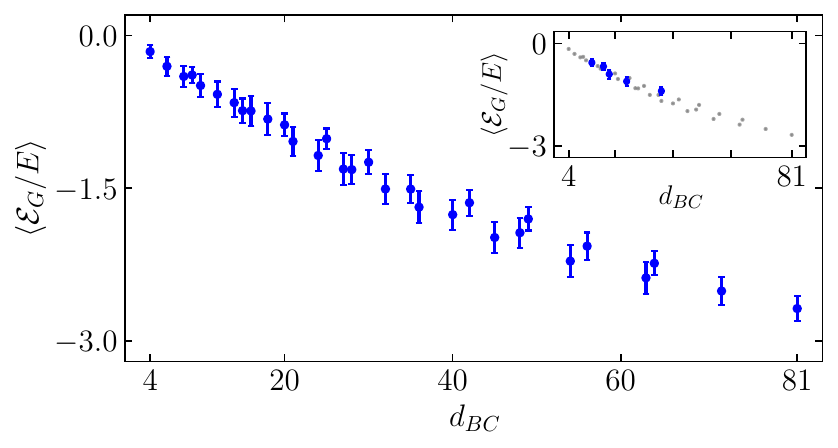}
    \caption{Ensemble-averaged dimensionless ergotropy gain, $\av{\erg_\G/E}$, for initially uncorrelated $B$ and $C$ plotted against $d_{BC}$. The plot comprises $45$ combinations of different dimensions $d_B = 2, \cdots, 9$; $d_C = d_B,\cdots, 9$. The ``error bars'' are given by the standard deviation of $\erg_\G/E$.
    For certain values of $d_{BC}$, there are more than one combinations of $d_B$ and $d_C$ such that $d_B d_C = d_{BC}$ (e.g., $12 = 2 \times 6 = 3 \times 4$). These ``duplicate'' configurations are not included in the main figure to avoid clutter. Rather, they are shown in the inset on the gray background of $\av{\erg_\G/E}$ for all $d_{BC}$.
    We see that, as $d_{BC}$ grows, uncorrelated states become increasingly lossy on average. In fact, the dependence of $\ln |\av{\erg_\G/E}|$ on $\ln d_{BC}$ is almost linear, which means that $\av{\erg_\G/E} \propto - N^\mu$. A linear fit of the $\ln |\av{\erg_\G/E}|$ vs $\ln d_{BC}$ curve suggests $\mu \approx 0.8$. Moreover, the standard deviation of $\erg_\G/E$ remains significantly smaller than $|\av{\erg_\G/E}|$, meaning that the likelihood of $\erg_\G > 0$ decreases with $d_{BC}$.
    }
    \label{fig:ProdState_AverageShift}
\end{figure}

Here we will demonstrate the high-dimensional behavior of $\erg_\G$ when the initial state of $BC$ is uncorrelated.

In Figs.~\ref{fig:rand_ProdSepGen}(a)--(c), we observed that, in this case, the ensemble bulk of $\erg_\G$ tends to become more and more negative as $d_{BC}$ grows. Here we study this phenomenon more systematically, by simulating all possible system dimensions such that $d_{BC} \leq 81$. Figure~\ref{fig:ProdState_AverageShift} shows $\av{\erg_\G}$---the ensemble average of $\erg_\G$---as a function of $d_{BC}$. Each dot comes with a vertical bar corresponding to $\mathrm{SD}(\erg_\G) = \sqrt{\av{\erg_G^2} - \av{\erg_\G}^2}$---the standard deviation of $\erg_\G$ over the ensemble. This plot clearly demonstrates that, generically, product states become increasingly lossy for ergotropy transport as the dimensions grow. This demonstrates that, while Lemma~\ref{thm:2qbt} does not hold in higher dimensions, correlations maintain their central role in increasing $\erg_G$ in that their lack generically leads to lossiness. 

Thus, in order to have lossless or gainful ergotropy transport in higher dimensions, one has two choices. One either (i) starts uncorrelated but has to perform highly controlled operations on finely tuned Hamiltonians or (ii) uses correlated initial states and gains some ``wiggle room'' in the degree of control and fine tuning (see also Appendix~\ref{app:reuse} for more on that). Given that the first option is unrealistic at high dimensions, especially if one deals with many-body systems, the second option remains as the only viable choice.

\section{Concentration of $\erg_\G$ and $\Delta I$ in higher dimensions}
\label{app:levy}

Here we will use the Levy's measure concentration bound \cite{Milman_book_2001} to show that the ergotropy gain and change of mutual information become concentrated in higher dimensions. As measure concentration bounds require Lipschitz continuity \cite{Milman_book_2001}, we will invoke the ergotropy continuity result proven in Ref.~\cite{Hovhannisyan_2024b}.

First, let us formulate Levy's inequality. Let $f(x)$ be a function mapping the $(N-1)$-dimensional unit sphere to the real line ($\mathbb{R}^1$). Suppose $f(x)$ is Lipschitz continuous with respect to the Euclidean metric in the $N$-dimensional flat space ($\mathbb{R}^N$) in which the sphere is embedded. Now, if $x$ is distributed according to the uniform Haar measure on the sphere, then \cite{Milman_book_2001}
\begin{align} \label{Levis}
    \prob\big[ |f(x) - \av{f}| > \ell \big] \leq 3 \, e^{- \ell^2 / \mathcal{B}_f^2},
\end{align}
where the averaging $\av{f}$ is over the Haar measure on the sphere and
\begin{align}
    \mathcal{B}_f = \frac{L_f}{\sqrt{\alpha N}},
\end{align}
with $L_f$ being the Lipschitz constant of $f$. The coefficient $\alpha$ is a universal constant; one can take $\alpha = 1 / (25 \pi)$ \cite{Watrous_book_2018} (Theorem 7.37). The quantity $\mathcal{B}_f$ controls the width of the distribution of $f$ around its average. Intuitively, it can be thought of as an analogue of the standard deviation of $f$.

Now, it was proven in Ref.~\cite{Hovhannisyan_2024b} that
\begin{align} \label{ergotropy_Lip}
    |\erg(\rho) - \erg(\sigma)| \leq 2 \opn{H} \DBu(\rho, \sigma),
\end{align}
where $\opn{H}$ is the operator norm of $H$, and in Refs.~\cite{Sekatski_2021, Hovhannisyan_2024b} that
\begin{align} \label{entropy_Lip}
    |S(\rho) - S(\sigma)| \leq L_S(d) \DBu(\rho, \sigma),
\end{align}
where
\begin{align}
    L_S(d) = \frac{2.322 \pi}{2 \ln 2} \ln d.
\end{align}

To see the implications of these facts on our problem, let us fix the Hamiltonians $H_B$ and $H_C$ and pick an energy-conserving unitary $U_{BC}$. Then, we can easily see that both $\erg_\G$ and $\Delta I$ are Lipschitz as functions of the initial state $\rho_{BC}$ with respect to the Bures metric.

\medskip

Indeed, starting with $\erg_G$ and using Eq.~\eqref{erggain}, we can write
\begin{align} \label{mammoth}
\begin{split}
    \big\vert \erg_\G(\rho_{BC}) - \erg_\G(\sigma_{BC}) \big\vert = \big\vert & \erg(\tilde{\rho}_B) + \erg(\tilde{\rho}_C) - \erg(\rho_B)
    \\
    &\!\!- \erg(\rho_C) - \erg(\tilde{\sigma}_B) - \erg(\tilde{\sigma}_C)
    \\
    &\!\!+ \erg(\sigma_B) + \erg(\sigma_C) \big\vert
    \\
    \leq \big\vert & \erg(\tilde{\rho}_B) - \erg(\tilde{\sigma}_B) \big\vert
    \\
    &\!\!+ \big\vert \erg(\tilde{\rho}_C) - \erg(\tilde{\sigma}_C) \big\vert
    \\
    &\!\!+ \big\vert \erg(\rho_B) - \erg(\sigma_B) \big\vert
    \\
    &\!\!+ \big\vert \erg(\rho_C) - \erg(\sigma_C) \big\vert
    \\
    \leq \phantom{\big\vert} & \!\! 2 \opn{H_B} \, \DBu(\tilde{\rho}_B, \tilde{\sigma}_B)
    \\
    &\!\!+ 2 \opn{H_B} \, \DBu(\rho_B, \sigma_B)
    \\
    &\!\!+ 2 \opn{H_C} \, \DBu(\tilde{\rho}_C, \tilde{\sigma}_C)
    \\
    &\!\!+ 2 \opn{H_C} \, \DBu(\rho_C, \sigma_C),
\end{split}
\end{align}
where in the last step we used Eq.~\eqref{ergotropy_Lip}.

Furthermore, invoking the unitary invariance and contractivity of the Bures metric, we have that
\begin{align} \label{trivium1}
\begin{split}
    \DBu(\tilde{\rho}_B, \tilde{\sigma}_B) &= \DBu(\tr_C [\tilde{\rho}_{BC}], \tr_C [\tilde{\sigma}_{BC}])
    \\
    &\leq \DBu(\tilde{\rho}_{BC}, \tilde{\sigma}_{BC})
    \\
    &= \DBu(\rho_{BC}, \sigma_{BC}),
\end{split}
\end{align}
where in the last step we took into account that $\tilde{\rho}_{BC} = U_{BC} \rho_{BC} U_{BC}^\dagger$ and $\tilde{\sigma}_{BC} = U_{BC} \sigma_{BC} U_{BC}^\dagger$. Similarly,
\begin{align} \label{trivium2}
\begin{split}
    \DBu(\rho_B, \sigma_B) &= \DBu(\tr_C [\rho_{BC}], \tr_C [\sigma_{BC}])
    \\
    &\leq \DBu(\rho_{BC}, \sigma_{BC}).
\end{split}
\end{align}
Through identical steps, we also establish that
\begin{align} \label{trivium3}
\begin{split}
    \DBu(\rho_C, \sigma_C) &\leq \DBu(\rho_{BC}, \sigma_{BC}),
    \\
    \DBu(\tilde{\rho}_C, \tilde{\sigma}_C) &\leq \DBu(\rho_{BC}, \sigma_{BC}).
\end{split}
\end{align}
All together, Eqs.~\eqref{mammoth}--\eqref{trivium3} mean that $\erg_\G$ is Lipschitz as a function of the initial state:
\begin{align}
    |\erg_\G(\rho_{BC}) - \erg_\G(\sigma_{BC})| \leq L_{\erg_\G} \, \DBu(\rho_{BC}, \sigma_{BC}),
\end{align}
with the Lipschitz constant
\begin{align} \label{koxac1}
    L_{\erg_\G} \leq 4 (\opn{H_B} + \opn{H_C}).
\end{align}

Lastly, let $\ket{\phi(\rho_{BC})}$ be the purification of $\rho_{BC}$ living in $\mathcal{H}_{BC} \otimes \mathcal{H}_{BC}$, where $\mathcal{H}_{BC}$ is the Hilbert space of $BC$. By definition, $\ket{\phi(\rho_{BC})}$ be the purification of $\rho_{BC}$ if $\rho_{BC} = \tr_{\mathrm{copy \; of} \; \mathcal{H}_{BC}} \big[\ket{\phi(\rho_{BC})} \bra{\phi(\rho_{BC})}\big]$. Then, it is easy to demonstrate (see, e.g., Ref.~\cite{Hovhannisyan_2024b}) that $\DBu(\rho_{BC}, \sigma_{BC}) \leq \Vert \ket{\phi(\rho_{BC})} - \ket{\phi(\sigma_{BC})} \! \Vert_2$, where $\Vert \cdot \Vert_2$ is the standard Euclidean metric in $\mathcal{H}_{BC} \otimes \mathcal{H}_{BC}$. With this, we also have that
\begin{align}
    |\erg_\G(\rho_{BC}) - \erg_\G(\sigma_{BC})| \leq L_{\erg_\G} \, \Vert \ket{\phi(\rho_{BC})} - \ket{\phi(\sigma_{BC})} \Vert_2.
\end{align}

\medskip

Turning to $\Delta I$ as a function of the initial state, we can upper-bound its Lipschitz constant by the exactly same reasoning as with $\erg_\G$. Indeed, using Eq.~\eqref{QMIchange} and Eq.~\eqref{entropy_Lip}, we write
\begin{align} \label{behemoth}
\begin{split}
    \big\vert \Delta I(\rho_{BC}) - \Delta I(\sigma_{BC}) \big\vert = \big\vert & S(\tilde{\rho}_B) + S(\tilde{\rho}_C) - S(\rho_B)
    \\
    &\!\!- S(\rho_C) - S(\tilde{\sigma}_B) - S(\tilde{\sigma}_C)
    \\
    &\!\!+ S(\sigma_B) + S(\sigma_C) \big\vert
    \\
    \leq \big\vert & S(\tilde{\rho}_B) - S(\tilde{\sigma}_B) \big\vert
    \\
    &\!\!+ \big\vert S(\tilde{\rho}_C) - S(\tilde{\sigma}_C) \big\vert
    \\
    &\!\!+ \big\vert S(\rho_B) - S(\sigma_B) \big\vert
    \\
    &\!\!+ \big\vert S(\rho_C) - S(\sigma_C) \big\vert
    \\
    \leq \phantom{\big\vert} & \!\! L_S(d_B) \DBu(\tilde{\rho}_B, \tilde{\sigma}_B)
    \\
    &\!\!+ L_S(d_B) \DBu(\rho_B, \sigma_B)
    \\
    &\!\!+ L_S(d_C) \DBu(\tilde{\rho}_C, \tilde{\sigma}_C)
    \\
    &\!\!+ L_S(d_C) \DBu(\rho_C, \sigma_C) \big].
\end{split}
\end{align}
Finally, using Eqs.~\eqref{trivium1}--\eqref{trivium3}, we conclude that $\Delta I$ is indeed Lipschitz as a function of the initial state:
\begin{align} \nonumber
    |\Delta I(\rho_{BC}) - \Delta I(\sigma_{BC})| &\leq L_{\Delta I} \, \DBu(\rho_{BC}, \sigma_{BC})
    \\
    & \leq L_{\Delta I} \, \Vert \ket{\phi(\rho_{BC})} - \ket{\phi(\sigma_{BC})} \Vert_2
\end{align}
with the Lipschitz constant
\begin{align} \label{koxac2}
    L_{\Delta I} \leq 2 [L_S(d_B) + L_S(d_C)] = \frac{2.322 \, \pi}{\ln 2} \ln d_{BC}.
\end{align}

\medskip

As discussed in Appendix~\ref{app:randstate}, the initial states of $BC$ are sampled from the Hilbert--Schmidt measure, which corresponds to sampling from the Fubini--Study measure on the pure states in $\mathcal{H}_{BC} \otimes \mathcal{H}_{BC}$. Now, let us pick an arbitrary basis $\{\ket{\zeta}\}_{\zeta = 0}^{d_{BC}^2 - 1}$ in $\mathcal{H}_{BC} \otimes \mathcal{H}_{BC}$. Each pure state $\ket{\phi}$ can be decomposed as $\ket{\phi} = \sum_{\zeta} c(\phi)_\zeta \ket{\zeta}$, with $c(\phi)_\zeta = \bra{\zeta} \! \phi\rangle$. The normalization condition yields
\begin{align*}
    \sum_{\zeta = 0}^{d_{BC}^2-1} \big( \mathrm{Re}[c(\phi)_\zeta]^2 + \mathrm{Im}[c(\phi)_{\zeta}]^2 \big) = 1,
\end{align*}
meaning that the set of pure states in $\mathcal{H}_{BC} \otimes \mathcal{H}_{BC}$ is equivalent to $\mathbb{S}^{2 d_{BC}^2 - 1}$---the $(2 d_{BC}^2 - 1)$-dimensional unit sphere embedded in $\mathbb{R}^{2 d_{BC}^2}$. Moreover, the Euclidean norm in $\mathcal{H}_{BC} \otimes \mathcal{H}_{BC}$ is equivalent to the Euclidean norm in $\mathbb{R}^{2 d_{BC}^2}$:
\begin{align*}
    \Vert \ket{\phi} - \ket{\psi} \Vert_2 = \left\Vert \overrightarrow{c(\phi)} - \overrightarrow{c(\psi)} \right\Vert_2,
\end{align*}
where
\begin{align*}
    \overrightarrow{c(\phi)} =
    \left[\begin{array}{c}
         \mathrm{Re}[c(\phi)_0]  \\
         \mathrm{Im}[c(\phi)_0] \\
         \vdots \\
         \mathrm{Re}[c(\phi)_{d_{BC}^2 - 1}] \\
         \mathrm{Im}[c(\phi)_{d_{BC}^2 - 1}]
    \end{array} \right].
\end{align*}
Most importantly, the Fubini--Study measure on the pure states in $\mathcal{H}_{BC} \otimes \mathcal{H}_{BC}$ is equivalent to the Haar measure on $\mathbb{S}^{2 d_{BC}^2 - 1}$ \cite{Bengtsson_book_2006}.

Putting all this together, we conclude that, when $\rho_{BC}$ is sampled from the Hilbert--Schmidt measure, the map $\mathcal{M}$: $\rho_{BC} \mapsto \overrightarrow{c(\phi(\rho_{BC}))}$ transforms $\erg_\G(\rho_{BC})$ into an $L_{\erg_\G}$-Lipschitz function on $\mathbb{S}^{2 d_{BC}^2 - 1}$, the argument of which is sampled uniformly, according to the Haar measure on $\mathbb{S}^{2 d_{BC}^2 - 1}$. Similarly, the map $\mathcal{M}$ transforms $\Delta I(\rho_{BC})$ to an $L_{\Delta I}$-Lipschitz function on $\mathbb{S}^{2 d_{BC}^2 - 1}$, the argument of which is sampled uniformly, according to the Haar measure on $\mathbb{S}^{2 d_{BC}^2 - 1}$. Applying Levy's inequality \eqref{Levis} to both of these functions, and keeping in mind Eqs.~\eqref{koxac1} and~\eqref{koxac2}, we thus find that
\begin{align}
    \prob\big[ | \erg_\G - \av{\erg_\G} | > \ell \big] \leq 3 \, e^{- \ell^2 / \mathcal{B}_{\erg_\G}^2} \, ,
\end{align}
with
\begin{align}
    \mathcal{B}_{\erg_\G} = \frac{\sqrt{25 \pi} L_{\erg_\G}}{\sqrt{2 d_{BC}^2}} \leq \frac{\sqrt{200 \pi} (\opn{H_B} + \opn{H_C})}{d_{BC}},
\end{align}
and
\begin{align} \label{DeltaI_conc}
    \prob\big[ |\Delta I - \langle \Delta I \rangle | > \ell \big] \leq 3 \, e^{- \ell^2 / \mathcal{B}_{\Delta I}^2} \, ,
\end{align}
with
\begin{align} \label{QMI-width}
    \mathcal{B}_{\Delta I} = \frac{\sqrt{25 \pi} L_{\Delta I}}{\sqrt{2 d_{BC}^2}} \leq \frac{65.951 \, \ln d_{BC}}{d_{BC}}.
\end{align}
Here the averages, $\av{\erg_\G}$ and $\av{\Delta I}$, are over the Hilbert--Schmidt measure in $\mathcal{H}_{BC}$. They are equal to the averages of the respective $\mathcal{M}$-transformed functions over the Haar measure on $\mathbb{S}^{2 d_{BC}^2 - 1}$.

In the simulations shown in Fig.~\ref{fig:Erg-MI_propeller}(b), we chose the dimensionless Hamiltonian to be such that $\opn{H} = O(d)$; see Appendix~\ref{app:randham}. More precisely, it is easy to check that, there, $\opn{H_B} + \opn{H_C} \leq 2 \max\{ d_B, d_C \}$. Thus, for the random Hamiltonian generation protocol of Appendix~\ref{app:randham}, that was used in Fig.~\ref{fig:Erg-MI_propeller}(b), 
\begin{align} \label{ErgG_conc}
    \prob\big[ | \erg_\G/E - \av{\erg_\G/E} | > \ell \big] \leq 3 \, e^{- \ell^2 / \widetilde{\mathcal{B}}_{\erg_\G}^2} \, ,
\end{align}
with
\begin{align} \label{dimless-H_ergH-width}
    \widetilde{\mathcal{B}}_{\erg_\G} < \frac{50.133 \, \max\{ d_B, d_C \}}{d_B d_C}.
\end{align}
As already mentioned in Appendix~\ref{app:randham}, there are many ways to choose a dimensionless Hamiltonian. In those cases where $B$ and $C$ are known to be many-body systems with short-range interactions, the norm of their Hamiltonians would be proportional to the number of particles in those systems, namely, $\opn{H_B} \propto \ln d_B$ and $\opn{H_C} \propto \ln d_C$. Thus, for such systems, we would have $\widetilde{\mathcal{B}}_{\erg_\G} = O \big( \frac{\ln d_{BC}}{d_{BC}} \big)$.

In any case, we see from Eqs.~\eqref{QMI-width} and~\eqref{dimless-H_ergH-width} that both distribution widths decay polynomially with $d_B$ and $d_C$, which is what we see in Fig.~\ref{fig:concentration}.

Lastly, by the very construction of the ensemble, for each $H_{BC}$, the process $\rho_{BC} \to \tilde{\rho}_{BC}$ is equally likely as the $\tilde{\rho}_{BC} \to \rho_{BC}$ process. Therefore, $\av{\erg_G} = 0$ and $\av{\Delta I} = 0$.

\section{Numerics for higher-dimensional systems}
\label{app:HighDim_Num}

\medskip

As proven in Appendix~\ref{app:levy}, when $\rho_{BC}$ is sampled from the Hilbert--Schmidt measure, $\erg_\G$ and $\Delta I$ concentrate around their ensemble-averages as $d_B$ and $d_C$ increase. This fact is also illustrated in Fig.~\ref{fig:Erg-MI_propeller}, where the ranges of $\erg_\G$ and $\Delta I$ in panel (b) are significantly narrower than in panel (a). Here we numerically confirm the concentration bounds in Appendix~\ref{app:levy}, and, on top of that, show that the relation between $\erg_\G$ and $\Delta I$ becomes more deterministic even if we factor out the concentration phenomenon.

\begin{figure}[!t]
    \centering
    \includegraphics[trim=0.2cm 0.2cm 0.1cm 0.2cm, clip, width=1.01\columnwidth]{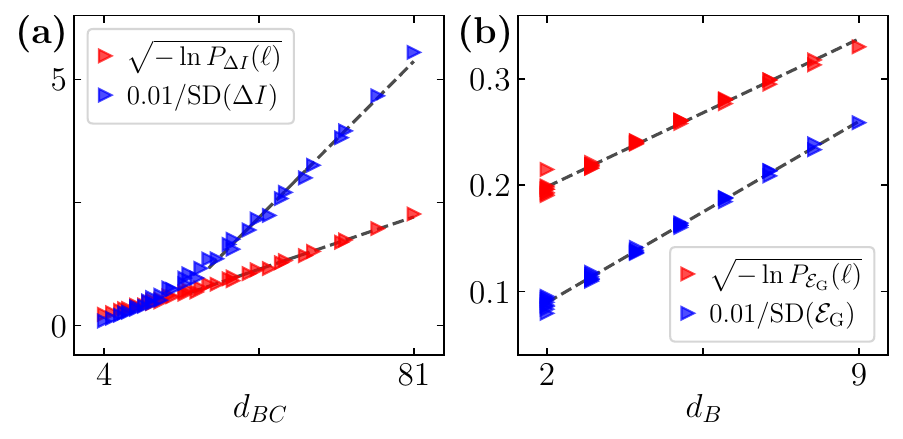}
    \caption{Here we plot the statistics of $\Delta I$ and $\erg_\G$ against $d_{BC}$ and $d_B$, in the sense of Eqs.~\eqref{skalka}. As in Fig.~\ref{fig:Erg-MI_propeller}, the states $\rho_{BC}$ are sampled from the Hilbert--Schmidt measure, and $H_{BC}$ and $U_{BC}$ are sampled according to the protocols described in Appendix~\ref{app:numerics}. The sample size for each dot is $10^6$ pints. Panel \textbf{(a)} is for $\Delta I$, and the red line shows that, for larger $d_{BC}$'s, $\sqrt{-\ln P_{\Delta I}(\ell)}$ scales linearly with $d_{BC}$. The blue line shows that $\mathrm{SD}(\Delta I)^{-1}$ scales as $d_{BC}$, both confirming and giving additional meaning to Eq.~\eqref{QMI-width}. Panel \textbf{(b)} does the same as Panel (a), but for $\erg_\G/E$ vs $d_B$, again, confirming Eqs.~\eqref{skalka} and~\eqref{dimless-H_ergH-width}. In both panels, $d_B \leq d_C$ run from $2$ to $9$ ($d_{BC} = 4, \cdots, 81$), $\ell = 0.005$, and the $0.01$ factor in front of $\mathrm{SD}(\Delta I)^{-1}$ and $\mathrm{SD}(\erg_\G)^{-1}$ is simply chosen to make the blue curves fit nicely with the red curves into the panels.}
    \label{fig:concentration}
\end{figure}

First, we confirm Eqs.~\eqref{DeltaI_conc}--\eqref{dimless-H_ergH-width}. Denoting
\begin{align}
    P_{\Delta I}(\ell) :=& \prob\big[ |\Delta I - \langle \Delta I \rangle | > \ell \big],
    \\
    P_{\erg_\G}(\ell) :=& \prob\big[ |\erg_\G/E - \langle \erg_\G/E \rangle | > \ell \big],
\end{align}
it is a simple exercise to check that, for large $d_{B}$, these bounds amount to
\begin{align} \label{skalka}
\begin{split}
    \sqrt{- \ln P_{\Delta I}(\ell)} &\propto \ell \frac{d_{BC}}{\ln d_{BC}},
    \\
    \sqrt{- \ln P_{\erg_\G}(\ell)} &\propto \ell \, d_B.
\end{split}
\end{align}
In Fig.~\ref{fig:concentration}, we plot $P_{\Delta I}(\ell)$ and $P_{\erg_\G}(\ell)$ against, respectively, $d_{BC}$ and $d_B$, and confirm that, at larger values of $d_B$ and $d_{BC}$, the dependence is indeed linear, which conforms with Eqs.~\eqref{skalka}; note that the presence of the logarithm becomes ``invisible'' for larger values of $d_{BC}$. Furthermore, given that $\mathcal{B}$ is somewhat similar to the standard deviation, one could expect that the standard deviations of $\Delta I$ and $\erg_\G/E$, $\mathrm{SD}(\Delta I)$ and $\mathrm{SD}(\erg_\G)$, would scale with $d_{BC}$ and $d_B$ as, respectively, $\mathcal{B}_{\Delta I}$ and $\mathcal{B}_{\erg_\G}$. And indeed, as Fig.~\ref{fig:concentration} shows, $\mathrm{SD}(\Delta I)^{-1}$ and $\mathrm{SD}(\erg_\G)^{-1}$ scale as, respectively, $d_{BC}$ and $d_B$. Figure~\ref{fig:concentration} thus corroborates the concentration bounds we obtained in Appendix~\ref{app:levy}, additionally showing that the relevant quantities scale with $d_B$ and $d_{BC}$ in the same way as the bounds. It also confirms the intuition that the standard deviations behave like $\mathcal{B}$'s.

\begin{figure}[!t]
    \centering
    \includegraphics[trim=0.2cm 0.2cm 0.2cm 0.2cm, clip, width=1.03\columnwidth]{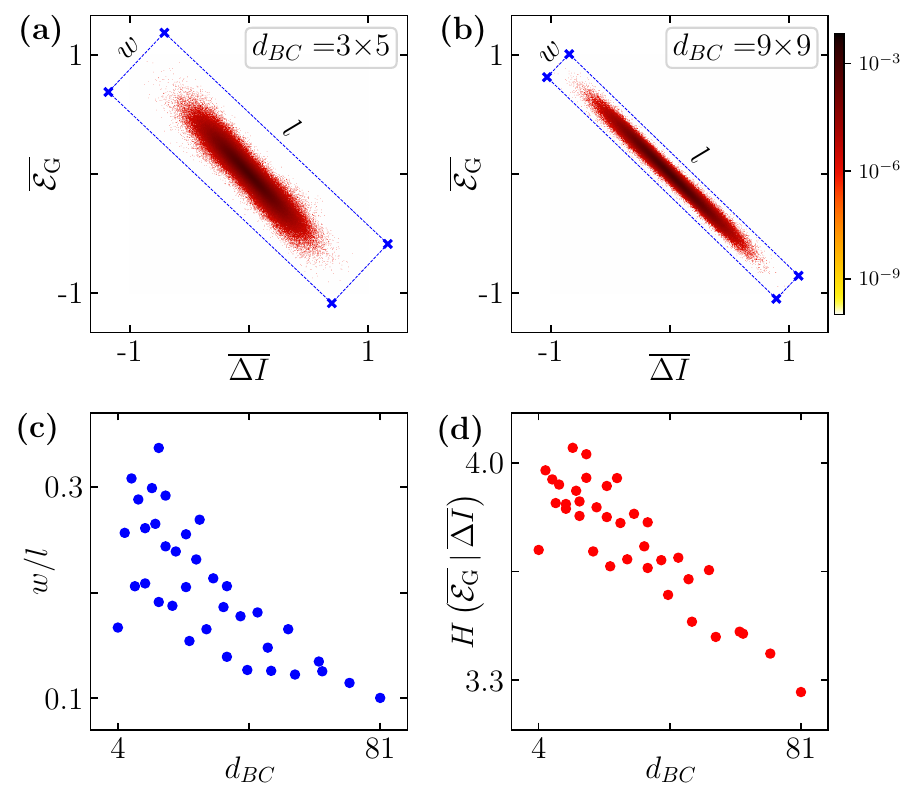}
    \caption{Here we demonstrate that the dispersion of the rescaled ergotropy gain, $\overline{\erg_\G}$ [see Eq.~\eqref{resc}], with respect to the rescaled mutual information change, $\overline{\Delta I}$ decreases at higher dimensions. Panels \textbf{(a)} shows an ensemble of $10^6$ tuples $(\overline{\Delta I}, \, \overline{\erg_\G})$ for $d_{BC} = 15$. Panel \textbf{(b)} shows the same, but for $d_{BC} = 81$. On each panel, the blue lines designate the rectangle with minimal area that contains all the points. The ratio $w/l$ of the rectangle's edges quantifies how disperse the dependence of $\overline{\erg_\G}$ on $\overline{\Delta I}$ is. Panel \textbf{(c)} shows $w/l$ for $4 \leq d_{BC} \leq 81$. We see that the ratio becomes small at large $d_{BC}$'s, meaning that $\overline{\erg_\G}$ becomes increasingly deterministic with respect to $\overline{\Delta I}$. Panel \textbf{(d)} shows the same trend, but now with the dispersion measured by $H\big( \overline{\erg_\G} \, \big\vert \overline{\Delta I} \, \big)$---the entropy of $\overline{\erg_\G}$ conditioned on $\overline{\Delta I}$. The smaller $H\big( \overline{\erg_\G} \, \big\vert \overline{\Delta I} \, \big)$, the less the dispersion of $\overline{\erg_\G}$ with respect to $\overline{\Delta I}$.
    }
    \label{fig:propeller_width}
\end{figure}

\medskip

The concentration bounds show that, as the dimensions increase, the relation between $\Delta I$ and $\erg_\G$ becomes more and more deterministic simply due to the fact that $\Delta I$ and $\erg_\G$ themselves become more deterministic as they concentrate around their respective averages. However, as we will argue here, this is not the whole story. Here we will argue that the relation between $\Delta I$ and $\erg_\G$ becomes \textit{even more deterministic} than is predicted by the concentration phenomenon.

To factor out the ``shrinkage'' due to concentration, we first pick an $N \gg 1$. Then, as in the simulations for Figs.~\ref{fig:Erg-MI_propeller} and~\ref{fig:concentration}, for each $d_B \times d_C$ configuration, we generate $N$ random tuples $(\rho_{BC}, H_{BC}, U_{BC})$ from which we obtain an ensemble of $N$ tuples $(\Delta I, \erg_\G/E)$. Now, for each $d_B \times d_C$, we take the observed $\max |\Delta I|$ and $\max |\erg_\G/E|$ over the corresponding ensemble and construct the rescaled ensemble consisting of $N$ tuples $(\overline{\Delta I}, \overline{\erg_G/E})$, where
\begin{align} \label{resc}
    \overline{\Delta I} = \frac{\Delta I}{\max |\Delta I|} \quad \mathrm{and} \quad \overline{\erg_\G} = \frac{\erg_\G}{\max |\erg_\G/E|}.
\end{align}
Thus, for all $d_B$ and $d_C$, $\overline{\Delta I} \in [-1, 1]$ and $\overline{\erg_\G} \in [-1, 1]$.

With the ranges of $\overline{\Delta I}$ and $\overline{\erg_\G}$ fixed, decrease in the dispersion in their relation with $d_{BC}$ cannot be attributed to measure concentration and must therefore be a separate phenomenon. And reduction of dispersion is indeed what we observe at higher $d_{BC}$'s. Indeed, for the same ensembles as in Fig.~\ref{fig:concentration} (where $N = 10^6$ and there are $45$ configurations of $d_B = 2, \cdots 9$ and $d_C = d_B, \cdots 9$), we made ``propeller'' plots of $(\overline{\Delta I}, \overline{\erg_\G})$. Two of these plots are shown in Panels (a) and (b) of Fig.~\ref{fig:propeller_width}. For each such plot, we first determine the convex hull of all the $N$ points (using the built-in ``ConvexHull'' function of the Python library SciPy). Then, we use the method called ``rotating calipers'' \cite{Toussaint_2000} to find the rectangle with the smallest area that contains the whole convex hull. The ratio of the width $w$ of the rectangle over its length $l$ [see Figs.~\ref{fig:propeller_width}(a) and (b)] is a measure of the dispersion of the dependence of $\overline{\erg_\G}$ on $\overline{\Delta I}$. The corresponding rectangles are shown in Figs.~\ref{fig:propeller_width}(a) and (b), and we see that for $d_{BC} = 81$ the rectangle is significantly narrower than for $d_{BC} = 15$. The dependence of $w/l$ on all values of $d_{BC}$ is shown in Fig.~\ref{fig:propeller_width}(c), where we see that $w/l$ tends to decrease as $d_{BC}$ becomes large.

Another way of measuring how disperse $\overline{\erg_\G}$ is with respect to $\overline{\Delta I}$, is to consider $H\big(\overline{\erg_\G} \, \big\vert \overline{\Delta I} \big)$---the entropy of the random variable $\overline{\erg_\G}$ conditioned on the random variable $\overline{\Delta I}$ (in this case, over our $N$-point ensemble). When $H(Y | X) = 0$, then $X$ is simply a deterministic function of $Y$, and the larger the $H(Y | X)$, the less deterministic the relation between $X$ and $Y$. For our ensembles, we calculate $H\big(\overline{\erg_\G} \, \big\vert \overline{\Delta I} \big)$ by coarse-graining the data into histograms and using the formula $H\big(\overline{\erg_\G} \, \big\vert \overline{\Delta I} \big) = H\big(\overline{\erg_\G},  \, \overline{\Delta I} \big) - H\big(\overline{\Delta I} \big)$, where $H(\cdot)$ is the Shannon entropy of the argument. The dependence of the resulting conditional entropy as a function of $d_{BC}$ is shown in Fig.~\ref{fig:propeller_width}(d). Here, too, we see a downward trend at high dimensions, once again showing that the relation between $\overline{\erg_\G}$ and $\overline{\Delta I}$ tends to become more deterministic.

Lastly, we note that our approach towards factoring out the concentration is not unique. Once could, for example, rescale the axes by the standard deviation, not by the maximal observed range. There also exist alternative entropy-based methods of assessing dispersion of one variable with respect to another.

\section{Bounds on the ergotropy gain for two qubits}
\label{app:propeller_bounds}

Here we refine the relationship between $\erg_\G$ and $\Delta I$ when $B$ and $C$ are qubits. We do so by establishing how much ergotropy gain is possible for a given change in mutual information.

First of all, let us show that the range of $\Delta I$ is $[-\ln 4, \ln 4]$ and that of $\erg_\G$ is $[-E, E]$. Indeed, from Eq.~\eqref{QMIchange} and the fact that, for a qubit, $0 \leq S(\rho) \leq \ln 2$, we immediately see that $-2 \ln 2 \leq \Delta I \leq 2 \ln 2$. Next, Eq.~\eqref{gain_vs_gap} and the fact that ergotropic gap is a nonnegative quantity imply that
\begin{align} \label{ergik}
-\tilde{\delta}^\erg_{BC} \leq \erg_\G \leq \delta^\erg_{BC}.
\end{align}
Now, for two identical qubits, Eq.~\eqref{loc_inv_erg_gap} yields
\begin{align} \label{klir}
    \delta_{BC}^\erg = E \big[\lambda^B_1 + \lambda^C_1 - \big( \lambda^{BC}_1 + \lambda^{BC}_2 - 2 \lambda^{BC}_3 \big)\big],
\end{align}
where
$\lambda^X_i$ are the eigenvalues of $\rho_X$ in decreasing order. And since $\lambda^B_1$ is the smaller eigenvalue of $\rho_B$, $\lambda^B_1 \leq 1/2$. Similarly, $\lambda^C_1 \leq 1/2$. Thus, by Eq.~\eqref{klir}, $\delta^\erg_{BC} \leq E$, and thus, by Eq.~\eqref{ergik}, $-E \leq \erg_G \leq E$.

\begin{figure}[!t]
	\centering
	\includegraphics[trim=0.2cm 0.3cm 0.1cm 0.1cm, clip, width=0.93\columnwidth]{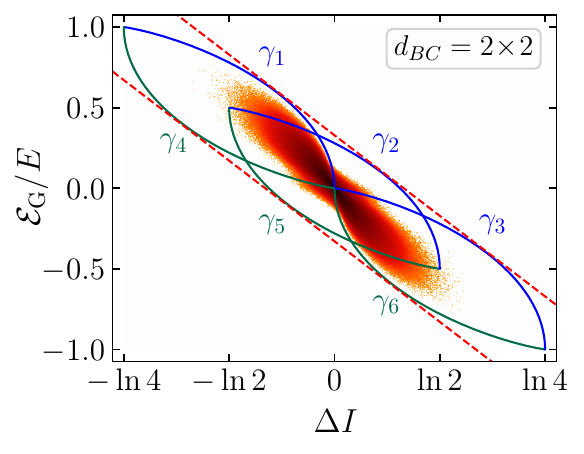}
	\caption{This is the exact copy of panel (a) of Fig.~\ref{fig:Erg-MI_propeller}, except that here $\gamma_1$--$\gamma_6$ are shown explicitly (blue and green lines) whereas $\gamma_u$ and $\gamma_d$ are omitted. The lines designated by $\gamma_1, \gamma_3, \gamma_4$ and $\gamma_6$ represent the most extreme cases achievable for two-qubit pure states. Whereas $\gamma_2$ and $\gamma_4$ give bounds for mixed states with two eigenvalues being $1/2$ and two eigenvalues being $0$. Note that while each $\gamma_i$ is a valid bound only in a limited domain of $\Delta I$, when appropriately combined [as per Eq.~\eqref{eq:lowerupperbound}], the resulting bounds $\gamma_u$ and $\gamma_d$ (see Fig.~\ref{fig:Erg-MI_propeller}) are exact and hold for the full range of $\Delta I$.}
	\label{fig:SM-propeller}
\end{figure}

\medskip

The main result of this section is that, by a combination of analytical derivations and numerical calculations (detailed below), we established that $\erg_\G$ is bounded as
\begin{align} \label{eq:lowerupperbound}
    \gamma_d(\Delta I) \leq \frac{\erg_\G}{E} \leq \gamma_u(\Delta I) 
\end{align}
where
\begin{align}
	\gamma_u(\Delta I) \! = \! \left\{ \! \begin{array}{ll}
		\gamma_1(\Delta I), & \; \Delta I \in [-\ln 4, -\ln 2],
		\\
		\max\{ \gamma_1(\Delta I),\gamma_2(\Delta I)\}, & \; \Delta I \in [-\ln 2, \, 0],
		\\
		\max\{ \gamma_2(\Delta I),\gamma_3(\Delta I)\}, & \; \Delta I \in [0, \, \ln 2],
		\\
		\gamma_3(\Delta I), & \; \Delta I \in [\ln 2, \, \ln 4],
	\end{array} \right.
\end{align}
and
\begin{align}
	\gamma_d(\Delta I)\! = \! \left\{ \! \begin{array}{ll}
		\gamma_4(\Delta I), & \; \Delta I \in [-\ln 4, -\ln 2],
		\\
		\min\{ \gamma_4(\Delta I),\gamma_5(\Delta I)\}, & \; \Delta I \in [-\ln 2, \, 0],
		\\
		\min\{ \gamma_5(\Delta I),\gamma_6(\Delta I)\}, & \; \Delta I \in [0, \, \ln 2],
		\\
		\gamma_6(\Delta I), & \; \Delta I \in [\ln 2, \, \ln 4].
	\end{array} \right.
\end{align}
The functions $\gamma_i$ setting the boundaries are defined as
\begin{align} \nonumber
    \gamma_1(x) &= 2 \, h_+^{-1}\!\left(\tfrac{\ln 4 + x}{2}\right) - 1, &&  x \in [-\ln 4, \, 0],
    \\ \label{eq:boundfunc-1}
    \gamma_2(x) &= 2 \, h_+^{-1}\!\left(\tfrac{\ln 2 + x}{2}\right) - \tfrac{3}{2}, && x \in [-\ln 2, \, \ln 2],
    \\ \nonumber
    \gamma_3(x) &= 2 \, h_+^{-1}\!\left(\tfrac{x}{2}\right) - 2, && x \in [0, \, \ln 4],
\end{align}
and
\begin{align} \nonumber
    \gamma_4(x) &= 2 - 2 \, h_+^{-1}\!\left(-\tfrac{x}{2}\right), && x \in [-\ln 4, \, 0],
    \\ \label{eq:boundfunc-2}
    \gamma_5(x) &= \tfrac{3}{2} - 2 \, h_+^{-1}\!\left(\tfrac{\ln 2 - x}{2}\right), && x \in [-\ln 2, \, \ln 2],
    \\ \nonumber
    \gamma_6(x) &= 1 - 2 \, h_+^{-1}\!\left(\tfrac{\ln 4 - x}{2}\right), && x \in [0, \, \ln 4].
\end{align}
Here
\begin{align*}
h_+^{-1} \; : \; [0, \, \ln 2] \;\; \mapsto \;\; [\tfrac{1}{2}, \, 1]
\end{align*}
is the monotonically decreasing branch of the inverse of the binary entropy function $h(x) = - x \ln x - (1-x) \ln (1-x)$, $x \in [0,1]$. The bounds $\gamma_1$--$\gamma_6$ are shown in Fig.~\ref{fig:SM-propeller}.

\medskip

To derive Eqs.~\eqref{eq:boundfunc-1} and~\eqref{eq:boundfunc-2}, we will use the solution to the quantum marginal problem for two qubits, and we will establish the veracity of the bounds in Eq.~\eqref{eq:lowerupperbound} numerically. The solution of the two-qubit marginal problem is summarized in Eqs.~\eqref{ineq1}--\eqref{ineq4}, and, for the largest local eigenvalues, it writes as
\begin{align} \label{eq:SolutionQuantumMarginal}
    \begin{split}
        \lambda^B_0 &\leq \lambda^{BC}_0 + \lambda^{BC}_1,
        \\
        \lambda^C_0 &\leq \lambda^{BC}_0 + \lambda^{BC}_1,
        \\
        \lambda^B_0 + \lambda^C_0 &\leq 2 \lambda^{BC}_0 + \lambda^{BC}_1 + \lambda^{BC}_2,
        \\
        \big\vert \lambda^B_0 - \lambda^C_0 \big\vert &\leq \min \big\{ \lambda^{BC}_0 - \lambda^{BC}_2, \lambda^{BC}_1 - \lambda^{BC}_3 \big\}.
\end{split}
\end{align}
To obtain the bounds, we will look at the two extreme cases of the last line in Eq.~\eqref{eq:SolutionQuantumMarginal}.
One extreme case are the pure states of $BC$ with $\spec(\rho_{BC}) = \{1,0,0,0\}$. The other extreme case are the mixed states of $BC$ such that $\spec(\rho_{BC}) = \{0.5,0.5,0,0\}$. The states with more mixed spectra---\{1/3, 1/3, 1/3, 0\} and \{1/4, 1/4, 1/4, 1/4\}---turn out to be uninformative when used as extreme cases, as they delimit narrow sets that do not bound the $\Delta I$--$\erg_\G$ relation.

We will first look at the pure case, i.e., when $\spec(\rho_{BC}) = \{1,0,0,0\}$. Here Eq.~\eqref{eq:SolutionQuantumMarginal} simply yields $\lambda^B_0 = \lambda^C_0 \leq 1$. And since $\tilde{\rho}_{BC}$ is also pure, $\tilde{\lambda}^B_0 = \tilde{\lambda}^C_0 \leq 1$ holds as well. Therefore,
\begin{align} \label{erggainpurestate} 
    \frac{\erg_\G}{2 E} &= \tilde{\lambda}^B_0 - \lambda^B_0,
    \\
    \frac{\Delta I}{2} &= h(\tilde{\lambda}^B_0) - h(\lambda^B_0),
\end{align}
which brings us to
\begin{align} \label{plor-1}
    \tilde{\lambda}^B_0 &= h_+^{-1}\left( \frac{\Delta I}{2} + h(\lambda^B_0)\right),
    \\ \label{plor-2}
    \lambda^B_0 &= h_+^{-1}\left( h(\tilde{\lambda}^B_0) - \frac{\Delta I}{2} \right).
\end{align}
Substituting Eq.~\eqref{plor-1} into Eq.~\eqref{erggainpurestate}, we get
\begin{align} \label{eq:EGofLambda}
    \frac{\erg_\G}{2 E} &= h_+^{-1}\left( \frac{\Delta I}{2} + h(\lambda^B_0)\right) - \lambda^B_0.
\end{align}
To find out how $\erg_\G$ changes with $\lambda^B_0$, let us look into its derivative with respect to it:
\begin{align} \label{flirv}
    \frac{\text{d}}{\text{d}\lambda^B_0} \frac{\erg_\G}{2 E} = \frac{h'(\lambda^B_0)}{h'\big(h_+^{-1}\big(\tfrac{\Delta I}{2} + h(\lambda^B_0)\big)\big)} - 1 .
\end{align}
Keeping in mind that $h_+^{-1}$ is monotonically decreasing, and that, in $[\tfrac{1}{2}, \, 1]$, $h$ and $h'$ are also monotonically decreasing, we have that
\begin{align*}
    h'\big(h_+^{-1}\big(\tfrac{\Delta I}{2} + h(\lambda^B_0)\big)\big) &\leq h'(\lambda^B_0) \leq 0, && \mathrm{for} \; \Delta I \leq 0,
    \\
    0 \geq h'\big(h_+^{-1}\big(\tfrac{\Delta I}{2} + h(\lambda^B_0)\big)\big) &\geq h'(\lambda^B_0), && \mathrm{for} \; \Delta I \geq 0.
\end{align*}
Thus, in view of Eq.~\eqref{flirv}, for $\Delta I \leq 0$, $\erg_\G$ decreases with $\lambda^B_0$, and therefore $\erg_\G(\lambda^B_0) \leq \erg_\G(\tfrac{1}{2})$. Whereas for $\Delta I \geq 0$, $\erg_\G$ increases with $\lambda^B_0$, meaning that $\erg_\G(\lambda^B_0) \leq \erg_\G(1)$. Reading from Eq.~\eqref{eq:EGofLambda}, we thus have
\begin{align} \label{Obergrenze-1}
\begin{split}
    \erg_\G/E &\leq \gamma_1(\Delta I), \quad \mathrm{for} \quad \Delta I \leq 0,
    \\
    \erg_\G/E &\leq \gamma_3(\Delta I), \quad \mathrm{for} \quad \Delta I \geq 0.
\end{split}
\end{align}
An analogous argumentation applied to Eq.~\eqref{plor-2} gives birth to the lower bounds
\begin{align} \label{Untergrenze-1}
\begin{split}
    \erg_\G/E &\geq \gamma_4(\Delta I), \quad \mathrm{for} \quad \Delta I \leq 0,
    \\
    \erg_\G/E &\geq \gamma_6(\Delta I), \quad \mathrm{for} \quad \Delta I \geq 0.
\end{split}
\end{align}
The functions $\gamma_i$ are defined in Eqs.~\eqref{eq:boundfunc-1} and~\eqref{eq:boundfunc-2}.

\medskip

Lets now discuss the case were $\spec(\rho_{BC}) = \{0.5,0.5,0,0\}$. Here Eqs.~\eqref{eq:SolutionQuantumMarginal} yield
\begin{align} \label{eq:MixedStateSolution-1}
    \lambda^B_0 + \lambda^C_0 &\leq 3/2,
    \\ \label{eq:MixedStateSolution-2}
    |\lambda^B_0 - \lambda^C_0| &\leq 1/2.
\end{align}
The last line gives two extreme cases: ($1$) $|\lambda^B_0 - \lambda^C_0| = 0$ (i.e., $\lambda^B_0 = \lambda^C_0$) or ($2$) $|\lambda^B_0 - \lambda^C_0| = 1/2$. In view of the fact that $\lambda_0$'s are the largest eigenvalues, case ($2$) simply means that either $\lambda^B_0 = \lambda^C_0 + 1/2$ or $\lambda^C_0 = \lambda^B_0 + 1/2$, which in turn implies that either $\lambda^B_0 = 1/2$ and $\lambda^C_0 = 1$ or $\lambda^B_0 = 1$ and $\lambda^C_0 = 1/2$. Since $U_{BC}$ preserves the spectrum of the global state, we have the same extreme cases for $\tilde{\lambda}^B_0$ and $\tilde{\lambda}^C_0$: ($\widetilde{1}$) $\tilde{\lambda}^B_0 = \tilde{\lambda}^C_0$ and ($\widetilde{2}$) either $\tilde{\lambda}^B_0 = 1$ and $\tilde{\lambda}^C_0 = 1/2$ or $\tilde{\lambda}^B_0 = 1/2$ and $\tilde{\lambda}^C_0 = 1$.

Now, cases ($1$)+($\widetilde{1}$) are already covered by the analysis above and yield the bounds in Eqs.~\eqref{Obergrenze-1} and~\eqref{Untergrenze-1}. Cases ($2$)+($\widetilde{2}$) simply yield $I(\rho_{BC}) = I(\tilde{\rho}_{BC}) = 0$ and $\erg(\rho_B) + \erg(\rho_C) = \erg(\tilde{\rho}_B) + \erg(\tilde{\rho}_C) = E$, which are trivial and uninformative. The remaining two combinations---cases ($1$)+($\widetilde{2}$) and ($2$)+($\widetilde{1}$)---provide two additional meaningful bounds. Let us take the ($1$)+($\widetilde{2}$) configuration. For it,
\begin{align} \label{klirs}
    \frac{\erg_\G}{E} &= 3/2 - 2 \lambda^B_0,
    \\ \label{klris}
    \Delta I &= \ln 2 - 2 \, h(\lambda^B_0).
\end{align}
As above, we express $\lambda^B_0$ through $\Delta I$ using Eq.~\eqref{klris} and substitute it into Eq.~\eqref{klirs}, which gives us $\erg_\G/E = \gamma_5(\Delta I)$. However, since, due to Eq.~\eqref{eq:MixedStateSolution-1}, $\lambda^B_0 \leq 3/4$, we have that $\erg_\G \geq 0$ and $-\ln 2 \leq \Delta I \leq -\tfrac{3}{2} \ln \tfrac{4}{3}$. Nonetheless, since the domain of $h^{-1}_+$ is $[0, \ln 2]$, the function $\gamma_5(\Delta I)$ is well-defined on the whole range of $[-\ln 2, \ln 2]$, and therefore we will take the full function
\begin{align}
    \erg_\G/E = \gamma_5(\Delta I), \quad \mathrm{for} \quad \Delta \in [-\ln 2, \, \ln 2]
\end{align}
as a boundary.

Analyzing the ($2$)+($\widetilde{1}$) configuration analogously, we arrive at the boundary function
\begin{align}
    \erg_\G/E = \gamma_2(\Delta I), \quad \mathrm{for} \quad \Delta \in [-\ln 2, \, \ln 2].
\end{align}
The functions $\gamma_2$ and $\gamma_5$ are defined in Eqs.~\eqref{eq:boundfunc-1} and~\eqref{eq:boundfunc-2}.

\medskip

Lastly, let us establish linear upper and lower bounds on $\erg_\G$ in terms of $\Delta I$. As the upper bound, we take the steepest tangential to $\gamma_1$ that does not cross $\gamma_2$ and $\gamma_3$. It turns out, that this line is also tangent to $\gamma_2$ and $\gamma_3$. Similarly, as a lower bound on $\erg_\G$, we take the steepest tangential to $\gamma_6$ that does not cross $\gamma_5$ and $\gamma_4$. This line, too, happens to be tangent to $\gamma_5$ and $\gamma_4$. Moreover, the two lines are parallel to each other and to the ``trend line'' connecting the most extreme points of $\erg_\G$. These points are $\erg_\G/E = 1, \Delta I = -\ln 4$ and $\erg_\G/E = -1, \Delta I = \ln 4$, and thus the trend line connecting them is
\begin{align}
    \erg_\G = -\frac{\Delta I}{\ln 4}.
\end{align}
Let us now find the value of $\Delta I$ for which the tangential of $\gamma_4$ is parallel to the trend line; namely, the solution of $d\gamma_4/d(\Delta I) = - 1/(\ln 4)$. This will give us the linear lower bound. Using the inverse function rule, we find that
\begin{align*}
    \frac{d\gamma_4}{d(\Delta I)} = \frac{1}{h'(h_+^{-1}(-\Delta I/2))} = \frac{1}{\ln \Big( \frac{1}{h_+^{-1}(-\Delta I/2)} - 1 \Big)},
\end{align*}
and equating it to $- 1/(\ln 4)$ immediately yields $\Delta I = -2 h (1/5)$. At this value of $\Delta I$, $\gamma_4$ evaluates to $2/5$, and therefore the line serving as the linear lower bound is
\begin{align*}
    - \frac{\Delta I}{\ln 4} - \left( \frac{h(1/5)}{\ln 2} - \frac{2}{5} \right).
\end{align*}
The above steps for $\gamma_5$ and $\gamma_6$ yield this same line, which proves that it is tangential to the three $\gamma$'s. Thus,
\begin{align} \label{lowbou}
    \frac{\erg_\G}{E} \geq - \frac{\Delta I}{\ln 4} - \left( \frac{h(1/5)}{\ln 2} - \frac{2}{5} \right).
\end{align}
A similar calculation with $\gamma_1$ (and then $\gamma_2$ and $\gamma_3$) yields the upper bound
\begin{align} \label{uppbou}
    \frac{\erg_\G}{E} \leq - \frac{\Delta I}{\ln 4} + \left( \frac{h(1/5)}{\ln 2} - \frac{2}{5} \right).
\end{align}
Using the lower bound in Eq.~\eqref{lowbou}, we immediately find the minimum value of $\Delta I$ necessary to ensure that $\erg_\G \geq 0$:
\begin{align}
    \Delta I_{\erg_\G \geq 0} \leq 2 \ln \frac{4}{5}.
\end{align}

\section{Reusing a correlated state for multiple transports}
\label{app:reuse}

Here we will calculate the ergotropy gain, and other relevant quantities, during iterations of the transport cycle. As detailed in the main text, the initial state is $\rho_{BC}^{(0)} = |\phi_{BC}^{(0)}\rangle \langle\phi_{BC}^{(0)}|$, with
\begin{align} \label{init_cohe}
    |\phi_{BC}^{(0)}\rangle = \cos \kappa \, \ket{0}_B \otimes \ket{0}_C + \sin \kappa \, \ket{1}_B \otimes \ket{1}_C,
\end{align}
and the transport unitary is
\begin{align*}
    U^{(\varepsilon)}_{BC} = \mathrm{SWAP} + 2 \nu^{(\varepsilon)}_{BC} \sin\frac{\varepsilon}{2} = \left( \begin{array}{cccc}
        1 & 0 & 0 & 0 \\
        0 & \sin \varepsilon & \cos \varepsilon & 0 \\
        0 & \cos \varepsilon & -\sin \varepsilon & 0 \\
        0 & 0 & 0 & 1 
    \end{array} \right).
\end{align*}
The ``error'' $\varepsilon \ll 1$ and ``correlation resource'' $0 < \kappa < \pi/4$. With such $\kappa$'s, the initial ergotropic gap is
\begin{align*}
    \delta^{(0)} = 2 E \sin^2 \kappa .
\end{align*}
The unitary operation performing a full cycle is
\begin{align}
\begin{split}
    U_\cyc^{(\varepsilon)} &= (\id_B \otimes U_\dra) \, U^{(\varepsilon)}_{BC} \, (U_\cha \otimes \id_C) \\
    &= \left( \begin{array}{cccc}
        \cos(\varepsilon) & 0 & 0 & \sin(\varepsilon) \\
        0 & 0 & 1 & 0 \\
        0 & 1 & 0 & 0 \\
        -\sin(\varepsilon) & 0 & 0 & \cos(\varepsilon) 
    \end{array} \right).
\end{split}
\end{align}
Hence, repeating the full transport $\iota$ times is equivalent to applying
\begin{align*}
    \big(U_\cyc^{(\varepsilon)}\big)^\iota = \left( \begin{array}{cccc}
        \cos(\iota\varepsilon) & 0 & 0 & \sin(\iota\varepsilon) \\
        0 & \frac{1+(-1)^\iota}{2} & \frac{1-(-1)^\iota}{2} & 0 \\
        0 & \frac{1-(-1)^\iota}{2} & \frac{1+(-1)^\iota}{2} & 0 \\
        -\sin(\iota\varepsilon) & 0 & 0 & \cos(\iota\varepsilon) 
    \end{array} \right)
\end{align*}
to $|\phi_{BC}^{(0)}\rangle$, which means that the state of $BC$ at the beginning of $(\iota + 1)$'th cycle is
\begin{align*}
    \big|\phi^{(\iota+1), \, \mathrm{in}}_{BC}\big\rangle =& \,\cos(\kappa - \iota\varepsilon) \ket{0}_B \otimes \ket{0}_C
    \\
    &+ \sin(\kappa - \iota\varepsilon) \ket{1}_B \otimes \ket{1}_C.
\end{align*}
In this state of $BC$, the reduced states of $B$ and $C$ are
\begin{align*}
\begin{split}
    \rho^{(\iota+1), \, \mathrm{in}}_B \! =& \ket{0}_B\!\bra{0} - \sigma_Z \sin^2(\kappa - \iota\varepsilon) ,
    \\
    \rho^{(\iota+1), \, \mathrm{in}}_C \! =& \ket{0}_C\!\bra{0} - \sigma_Z \sin^2(\kappa - \iota\varepsilon),
\end{split}
\end{align*}
where $\sigma_Z$ is the Pauli $Z$ matrix. As long as
\begin{align} \label{inj_pos_cond}
    \iota \leq \frac{\kappa}{\varepsilon} + \frac{\pi}{4 \varepsilon},
\end{align}
both $\rho^{(\iota+1), \, \mathrm{in}}_B$ and $\rho^{(\iota+1), \, \mathrm{in}}_C$ are passive. Thus, once $U_\cha$ is applied on $B$,
\begin{align} \label{pauk}
    \erg^{(\iota + 1)}_{\mathrm{injected \; via \;} B} = E \cos(2\kappa - 2\iota\varepsilon)
\end{align}
amount of ergotropy is injected into the system via $B$, and the pre-transport state $BC$ becomes
\begin{align*}
    \big|\phi^{(\iota+1)}_{BC}\big\rangle =& \, \cos(\kappa - \iota\varepsilon) \ket{1}_B \otimes \ket{0}_C
    \\
    &+ \sin(\kappa - \iota\varepsilon) \ket{0}_B \otimes \ket{1}_C.
\end{align*}
So, the pre-transport reduced states are
\begin{align}
\begin{split}
    \rho^{(\iota+1)}_B \! =& \ket{0}_B\!\bra{0} - \sigma_Z \cos^2(\kappa - \iota\varepsilon),
    \\
    \rho^{(\iota+1)}_C \! =& \ket{0}_C\!\bra{0} - \sigma_Z \sin^2(\kappa - \iota\varepsilon),
\end{split}
\end{align}
and
\begin{align} \label{begEBC}
\begin{split}
    \erg\big(\rho_B^{(\iota+1)}\big) &= E \cos(2\kappa - 2\iota\varepsilon),
    \\
    \erg\big(\rho_C^{(\iota+1)}\big) &= 0.
\end{split}
\end{align}
Next, the application of the transport unitary $U^{(\varepsilon)}_{BC}$ leaves the system in the state
\begin{align*}
    \big|\tilde{\phi}^{(\iota+1)}_{BC}\big\rangle = & \cos(\kappa - (\iota+1)\varepsilon) \ket{0}_B \otimes \ket{1}_C
    \\
    &+ \sin(\kappa - (\iota+1)\varepsilon) \ket{1}_B \otimes \ket{0}_C,
\end{align*}
in which
\begin{align}
\begin{split}
    \tilde{\rho}^{(\iota+1)}_B \! =& \ket{0}_B\!\bra{0} - \sigma_Z \sin^2(\kappa - (\iota+1)\varepsilon),
    \\
    \tilde{\rho}^{(\iota+1)}_C \! =& \ket{0}_C\!\bra{0} - \sigma_Z \cos^2(\kappa - (\iota + 1)\varepsilon).
\end{split}
\end{align}
It is straightforward to check that, whenever
\begin{align} \label{ext_pos_cond}
    \iota + 1 \leq \frac{\kappa}{\varepsilon} + \frac{\pi}{4 \varepsilon},
\end{align}
the local ergotropies are
\begin{align} \label{endEBC}
\begin{split}
    \erg\big(\tilde{\rho}_B^{(\iota+1)}\big) &= 0,
    \\
    \erg\big(\tilde{\rho}_C^{(\iota+1)}\big) &= E \cos(2\kappa - 2(\iota + 1)\varepsilon),
\end{split}
\end{align}
and the optimal ergotropy-extracting unitary for $\tilde{\rho}_C^{(\iota+1)}$ is simply $\sigma_X$. Thus, the $U_\dra$ applied at the end of the cycle fully drains $C$, extracting
\begin{align} \label{erg_ext_iota}
    \erg^{(\iota + 1)}_{\mathrm{extracted \; via \;} C} = \erg\big(\tilde{\rho}_C^{(\iota+1)}\big) = E \cos(2\kappa - 2(\iota + 1)\varepsilon).
\end{align}
We can now read off the ergotropy gain during the $(\iota + 1)$'th cycle directly from Eqs.~\eqref{begEBC} and~\eqref{endEBC}:
\begin{align}
    \erg_\G^{(\iota + 1)} = 2 E \sin(2 \kappa - 2 \iota\varepsilon - \varepsilon) \sin \varepsilon,
\end{align}
which coincides with Eq.~\eqref{ergGiota}. From this expression, we immediately see that
\begin{align} \label{ergG_pos_cond}
    \erg_G^{(\iota)} \geq 0 \quad \mathrm{for} \quad 1 \leq \iota \leq \frac{\kappa}{\varepsilon} + \frac{1}{2}.
\end{align}
Note that the $\iota$'s in Eq.~\eqref{ergG_pos_cond} automatically satisfy the conditions in Eqs.~\eqref{inj_pos_cond} and~\eqref{ext_pos_cond}. But the ranges of $\iota$ in Eqs.~\eqref{inj_pos_cond} and~\eqref{ext_pos_cond} go much beyond the range in Eq.~\eqref{ergG_pos_cond}, meaning that, up to some finite value of $\iota$, $BC$ can maintain a positive throughput, namely, $\erg^{(\iota)}_{\mathrm{extracted \; via \;} C} \geq 0$, even when $\erg_G^{(\iota)} \leq 0$.

Finally, from Eqs.~\eqref{erg_ext_iota} and~\eqref{begEBC}, we can write the total amount of ergotropy losslessly transported through $BC$ [defined in Eq.~\eqref{erg_ext_total}] explicitly:
\begin{align} \label{fuet}
    \ergt^+ = E \sum_{\iota = 1}^{\left\lfloor \frac{\kappa}{\varepsilon} + \frac{1}{2} \right\rfloor} \cos(2 \kappa - 2 \iota \varepsilon).
\end{align}

For the values chosen in the main text, namely, $\kappa = \pi/8$ and $\epsilon = 0.03$, Eq.~\eqref{ergG_pos_cond} tells us that the transport will be lossless (moreover, gainful) during the first $13$ iterations, and Eq.~\eqref{fuet} yields $\ergt^+ \approx 11.84 E$. Using Eq.~\eqref{pauk}, it is easy to calculate that the total amount of ergotropy that is injected into $B$ during these $13$ iterations will be $\approx 11.54 E < \ergt^+$.

\section{Ergotropy transport and quantum marginal problem}
\label{app:QuanMargProb}

In this section, we will show that our ergotropy transport problem is intimately connected to the problem of the compatibility of the spectra of local density matrices with the spectrum of the global density matrix, which is known as the ``quantum marginal problem'' (QMP) \cite{Klyachko_2006, Christandl_2005}.

For that, we first point out that, since the transport-performing operation $U_{BC}$ is energy conserving [see Eq.~\eqref{erg_con_uni}], $\rho_{BC}^\downarrow = \tilde{\rho}_{BC}^\downarrow$. Hence, due to Eq.~\eqref{loc_inv_erg_gap}, the ergotropy gain in arbitrary Hilbert-space dimensions can be written as
\begin{align} \label{erggain_general}
    \erg_{G} = \sum_{k=0}^{d_B-1}  E^B_k \big( \lambda^B_k - \tilde{\lambda}^B_k \big) + \sum_{k=0}^{d_C-1} E^C_k \big(\lambda^C_k - \tilde{\lambda}^C_k \big).
\end{align}
Here $E^X_k$ ($X = B, C$) are the energy levels of the local Hamiltonian $H_X$ ordered increasingly ($E^X_0 \leq \cdots \leq E^X_{d_X}$) and $\lambda^X_k$ are the decreasingly ordered eigenvalues of the reduced density matrix $\rho_X$ ($\lambda^X_0 \geq \cdots \geq \lambda^X_{d_X}$); the same applies to post-transport states $\tilde{\rho}_X$ and their eigenvalues $\tilde{\lambda}^X_k$.

Thus, the lossiness (or gainfulness) of ergotropy transport over an energy-conserving channel is determined solely by the initial and final \textit{spectra} of the local states. Importantly, since transport is realized by a global unitary operation ($U_{BC}$), we have that $\spec(\rho_{BC}) = \spec(\tilde{\rho}_{BC})$. Therefore, the initial pair of local spectra ($\{ \lambda^B_k \}_k$ and $\{ \lambda^C_k \}_k$) and the final pair of local spectra ($\{ \tilde{\lambda}^B_k \}_k$ and $\{ \tilde{\lambda}^C_k \}_k$) must be compatible with the \textit{same} global spectrum. So, both local spectra are related to each other by the fact that they must be a solution to the QMP for a given global density matrix spectrum.

This relation constitutes the connection of the QMP to the ergotropy transport problem. Below, to show the strength of this connection, we prove Lemma~\ref{thm:2qbt} using the solution of the two-qubit QMP \cite{Bravyi_2003}. Furthermore, we show that the connection between the two problems goes both ways, by showing that a large subset of QMP inequalities derived in Ref.~\cite{Klyachko_2006} can be obtained from the simple fact that the ergotropic gap is nonnegative.

\subsection{From QMP to Lemma~2}

As discussed in Appendix~\ref{app:two_qubits}, for two qubits, we can set $E^B_0 = E^C_0 = 0$ without loss of generality, and to have nontrivial global-energy-conserving transport, $E^B_1 = E^C_1 = E$ must hold. In this energetic configuration, Eq.~\eqref{erggain_general} reduces to
\begin{align} \label{erggain_twoqubits}
    \erg_\G = E \big( \lambda^B_1 + \lambda^C_1 - \tilde{\lambda}^B_1 - \tilde{\lambda}^C_1 \big).
\end{align}
Thus, lossy transport (i.e., $\erg_\G \leq 0$) is equivalent to
\begin{align} \label{flir}
    \lambda^B_1 + \lambda^C_1 \leq \tilde{\lambda}^B_1 + \tilde{\lambda}^C_1.
\end{align}


Now, the solution of the QMP for two qubits (say, $B$ and $C$) \cite{Bravyi_2003} is as follows. The local states $\rho_B$ and $\rho_C$ with spectra $\big\{ \lambda^B_0 \geq \lambda^B_1 \big\}$ and $\big\{ \lambda^C_0 \geq \lambda^C_1 \big\}$ can be the marginals of a global state $\rho_{BC}$ with the spectrum $\big\{\lambda^{BC}_0 \geq \lambda^{BC}_1 \geq \lambda^{BC}_2 \geq \lambda^{BC}_3\big \}$ only if
\begin{align} \label{ineq1}
    \lambda^B_1 &\geq \lambda^{BC}_2 + \lambda^{BC}_3,
    \\ \label{ineq2}
    \lambda^C_1 &\geq \lambda^{BC}_2 + \lambda^{BC}_3,
    \\ \label{ineq3}
    \lambda^B_1 + \lambda^C_1 &\geq \lambda^{BC}_1 + \lambda^{BC}_2 + 2 \lambda^{BC}_3,
    \\ \label{ineq4}
    \big\vert \lambda^B_1 - \lambda^C_1 \big\vert &\leq \min \big\{ \lambda^{BC}_0 - \lambda^{BC}_2, \, \lambda^{BC}_1 - \lambda^{BC}_3 \big\} .
\end{align}

In the setting of Lemma~\ref{thm:2qbt}, $\rho_{BC} = \rho_B \otimes \rho_C$, and therefore
\begin{align} \label{specAB}
\begin{split}
    \spec(\rho_{BC}) &= \spec(\tilde{\rho}_{BC})
    \\
    =& \{ \lambda^B_0 \lambda^C_0, \, \lambda^B_0 \lambda^C_1, \, \lambda^B_1 \lambda^C_0, \, \lambda^B_1 \lambda^C_1\}.
\end{split}
\end{align}
We immediately see that, with the above ordering convention, $\lambda^{BC}_0 = \lambda^B_0 \lambda^C_0$ and $\lambda^{BC}_3 = \lambda^B_1 \lambda^C_1$. For $\lambda^{BC}_1$ and $\lambda^{BC}_2$, we have two options: either $\lambda^{BC}_1 = \lambda^B_0 \lambda^C_1$ and $\lambda^{BC}_2 = \lambda^B_1 \lambda^C_0$ or $\lambda^{BC}_1 = \lambda^B_1 \lambda^C_0$ and $\lambda^{BC}_2 = \lambda^B_0 \lambda^C_1$. In any case, we have that $\lambda^{BC}_1 + \lambda^{BC}_2 = \lambda^B_0 \lambda^C_1 + \lambda^B_1 \lambda^C_0$. Therefore, applying Eq.~\eqref{ineq3} to post-transport spectra $\big\{ \tilde{\lambda}^B_0 \geq \tilde{\lambda}^B_1 \big\}$ and $\big\{ \tilde{\lambda}^C_0 \geq \tilde{\lambda}^C_1 \big\}$, we get
\begin{align}
    \begin{split}
        \tilde{\lambda}^B_1 + \tilde{\lambda}^C_1 &\geq \lambda^B_0 \lambda^C_1 + \lambda^B_1 \lambda^C_0 + 2 \lambda^B_1 \lambda^C_1
        \\
        &= \lambda^B_1 + \lambda^C_1,
    \end{split}
\end{align}
which proves Eq.~\eqref{flir} and thus Lemma~\ref{thm:2qbt}.

\subsection{From nonnegative ergotropic gap to QMP}
\label{app:QuanMargProb_reverse}

Let us now show that we can turn around part of the argumentation used in the previous subsection to provide a simple proof for a general family of inequalities for the QMP that were previously obtained in Ref.~\cite{Klyachko_2006} using rather sophisticated mathematical machinery.

Suppose we have a bipartite system $BC$ such that $\spec(\rho_B) = \big\{ \lambda^B_0 \geq \cdots \geq \lambda^B_{d_B} \big\}$, $\spec(\rho_C) = \big\{ \lambda^C_0 \geq \cdots \geq \lambda^C_{d_C} \big\}$, and $\spec(\rho_{BC}) = \big\{ \lambda^{BC}_0 \geq \cdots \geq \lambda^{BC}_{d_{BC}} \big\}$.

Now, for \textit{any} Hamiltonian $H_{BC} = H_B \otimes \id_C + \id_B \otimes H_C$, we have that $\delta_{BC}^\erg \geq 0$. To write this explicitly, following the notation used in Eq.~\eqref{erggain_general}, let $E^B_k$ and $E^C_k$ be the increasingly-ordered eigenvalues of, respectively, $H_B$ and $H_C$. Also, let $(E^B + E^C)_k$ be the increasingly-ordered eigenvalues of $H_{BC}$ (which is simply the set $\{ E^B_k + E^C_j \}_{k=0}^{d_B-1}\phantom{|}_{j=0}^{d_C-1}$, but reordered from smallest to largest). In this notation,
\begin{align}
\begin{split}
    \delta^\erg_{BC} =& \sum_{k=0}^{d_B-1} E^B_k \lambda^B_k + \sum_{k=0}^{d_C-1} E^C_k \lambda^C_k
    \\
    &- \sum_{k=0}^{d_B d_C - 1} (E^B + E^C)_k \lambda^{BC}_k,
\end{split}
\end{align}
and the fact that $\delta_{BC}^\erg \geq 0$, leads us to
\begin{align} \label{klyacha}
    \sum_{k=0}^{d_B-1} E^B_k \lambda^B_k + \sum_{k=0}^{d_C-1} E^C_k \lambda^C_k \geq \sum_{k=0}^{d_B d_C - 1} (E^B + E^C)_k \lambda^{BC}_k.
\end{align}
We emphasize that this inequality holds for \textit{any} choice of $\big\{ E^B_k \big\}_k$ and $\big\{ E^C_k \big\}_k$. Since the choice of the ``zero level'' of energy is arbitrary, we can, without loss of generality, additionally set
\begin{align*}
    \sum_{k=0}^{d_B - 1} E^B_k = \sum_{k=0}^{d_C-1} E^C_k = 0.
\end{align*}

The family of inequalities given by Eq.~\eqref{klyacha} coincides with the subset of QMP inequalities in ``Example 3.5.1'' in Ref.~\cite{Klyachko_2006}. While this family is strictly smaller than the full set of QMP inequalities (Theorem 3.4.1 in Ref.~\cite{Klyachko_2006}), it does cover a lot of ground, especially in low dimensions. To see that, let us go to the case of two qubits. By choosing $H_B = E \ket{1}_B \!\bra{1}$ and $H_C = \nul$, we obtain Eq.~\eqref{ineq1}. By choosing $H_B = \nul$ and $H_C = E \ket{1}_C \!\bra{1}$, we obtain Eq.~\eqref{ineq2}. And by choosing $H_B = E \ket{1}_B \!\bra{1}$ and $H_C = E \ket{1}_C \!\bra{1}$, we obtain Eq.~\eqref{ineq3}. So, three out of four QMP inequalities for two qubits can be obtained from the nonnegativity of the ergotropic gap.

\end{document}